\documentclass[onecolumn,pre,preprint,amsmath,amssymb,superscriptaddress]{revtex4-1}

\usepackage{amsmath}
\usepackage{bm}
\usepackage[dvipdfmx]{graphicx}
\usepackage{mathrsfs}
\usepackage{color}

\begin{document}

\title{Interfacial friction dictates long-range force propagation in tissues}
\date{\today}

\author{Yuting Lou}
\email {chelinqueen@hotmail.com}
\address {Mechanobiology Institute, National University of Singapore, Singapore 117411, Singapore}
\author{Takumi Kawaue}
\address {Mechanobiology Institute, National University of Singapore, Singapore 117411, Singapore}
\author{Ivan Yow}
\address {Mechanobiology Institute, National University of Singapore, Singapore 117411, Singapore}
\author{Yusuke Toyama} 
\address {Mechanobiology Institute, National University of Singapore, Singapore 117411, Singapore}
\author{Jacques Prost}
\address {Mechanobiology Institute, National University of Singapore, Singapore 117411, Singapore}
\address {Laboratoire Physico Chimie Curie, Institut Curie, Paris Science et Lettres Research University, CNRS UMR168, 75005 Paris, France}
\author{Tetsuya Hiraiwa}
\email {mbithi@nus.edu.sg}
\address {Mechanobiology Institute, National University of Singapore, Singapore 117411, Singapore}

\begin{abstract}
Tissues are characterized by layers of functional units such as cells and extracellular matrix (ECM). 
Nevertheless, how dynamics at interlayer interfaces help transmit cellular forces in tissues remains overlooked. Here, we investigate a multi-layer system where a layer of epithelial cells is seeded upon an elastic substrate in contact with a hard surface. Our experiments show that, upon a cell extrusion event in the cellular layer, long-range wave propagation emerges in the substrate only when the two substrate layers were weakly attached to each other. %to allow interfacial sliding between them. 
We then derive a theoretical model which quantitatively reproduces the wave dynamics and explains how frictional sliding between substrate layers helps propagate cellular forces at a variety of scales, depending on the stiffness, thickness, and slipperiness of the substrate. These results highlight the importance of interfacial friction between layers in transmitting mechanical cues in tissues \textit{in vivo}.
\end{abstract}

%\pacs{...}

\maketitle

\section*{Introduction}

It is now widely accepted that mechanics is a key player for  understanding how cells coordinate tissue morphodynamics and homeostasis.
Cells can sense the mechanical stress or strain in a tissue or in the microenvironment, which can regulate cell migration and spreading, cell division rate and differentiation \cite{De2007,PhysRevE.78.031923,Hoffman2011,Iwadate2013,Cui2015,Gudipaty2017,Asano2018,roca2017quantifying,Merkel2007,maloney2008influence,Sen2009,Buxboim297}. %Previous studies paid much attention on the static mechanical responses of cells to ECM. For instance, there are several earlier works investigating to what depth cells can feel ECM and how cells can sense different stiffness of ECM.
%Intercellular force 
Simultaneously, cells can generate mechanical stresses and forces in a tissue. Earlier works revealed multiple ways in which forces originate in {\it in situ} biological systems, like anisotropic and pulsed actomyosin contractions \cite{Bertet2004,Martin2009,Heisenberg2013} and apoptotic forces \cite{Toyama1683,Teng2011}, and 
%the relevance for the 
their relevance for
morphogenetic dynamics.
Such mechanical stresses or forces generated by cells in a tissue can function as a mechanical cue for the autonomous coordination of cellular behaviors within the tissue. For example, the spatiotemporal pattern of intercellular forces in a tissue can steer collective cell migration for wound healing \cite{Trepat2009,Tambe2011,Aoki2017,Hino2020,boocock2021theory,Fukuyama2020arXiv}.
%temporal mechanical patterns around a dividing cell
Cell cycle progression is also regulated by intercellular force \cite{Uroz2018}, suggesting that the force may be relevant for maintaining tissue homeostasis.

% state-of-the-arts%factors which can influence such autonomous coordination by a mechanical cue may be the way the mechanical stress or the force exerted by a cell is transmitted to other cells. 
%Extracellular matrix (ECM), as well as cells themselves, may work as an important mediator of mechanical cues. %The mechanical response of ECM to the force exerted by cells has been studied for certain situations. For instance, several groups have addressed how deeply cells can feel ECM, both experimentally using various thickness of substrates and theoretically solving the elastic substrate model \cite{Merkel2007,maloney2008influence,Sen2009,Buxboim297}. These studies assumed the simplest situation, in which the bottom of the substrate underlying the tissue strongly adheres to another rigid base so that the substrate cannot move. Owing to this immobility, the deformation of a substrate is constrained by the substrate thickness and therefore the stress cannot propagate farther than the scale of the thickness. However, in a multicellular organism, it is likely that tissues are not always strongly bonded to some rigid ECM structures because of the complexity in ECM constitutions as well as remodeling mechanisms, which may allow ECM to mediate forces at long distances. 

One important fact to consider is that tissue systems are composed of multiple layers of diverse components including cells and extracellular matrix (ECM). This raises a  question how such layered structure affects the way of force transmission in tissues. Previous studies paid much attention on static features; e.g. static deformation profile of a thin gel with a fixed base under a cell has been clarified by elastic gel theory combined with quantitative biomechanical experiments \cite{Merkel2007,maloney2008influence,Sen2009,Buxboim297}. However, in a tissue composed of multiple layers, there can be sliding between layers under forces, which is indeed a subject studied in the field of (bio-)tribology for bioengineering purpose \cite{zhou2015biotribology}. The classical example of this is the bone tissue system, where frictional contact occurs at bone-tendon, bone-muscle or cartilage-cartilage interfaces \cite{barnett1962lubrication,shacham2010measurements,merkher2006rational,katta2008biotribology}. Nevertheless, the effect of tribological characteristics such as layer sliding on force propagation remains overlooked in the study of tissue mechanics and so far no experiment has been developed to study this aspect in a controlled way. 

In this article, we aim to study the role of layer sliding in transmitting the forces in tissues to understand its biological significance. We approach this issue by focusing on a simplified three-layered system \textit{in vitro}, where an epithelial tissue layer locates on an elastic gel substrate which adheres to a glass base. A force origin is induced in the tissue layer by triggering an apoptotic event, and we study the force propagation with and without the sliding at the gel-glass surface. A well-defined concentric propagation of gel deformation over several cell diameters can be observed only when the gel-glass sliding occurs. To study the physical mechanisms underlying this novel form of long-range propagation, we apply the elastic gel theory in which the tissue responses is decoupled from the gel mechanics and analyze the gel deformation dynamics under various assumptions of tissue stress field as an external force source exerted to the gel surface. The theory reveals that the competition between the elasticity of a gel and gel-glass friction can quantitatively reproduce this long-range propagation of gel deformation observed in our experiments and explains why this long-range propagation of substrate deformation occurs only on layers with weakly adhesive bases. Analytical solutions predict that the scale of this propagation will change drastically when varying the mechanical properties of the substrate layer within physiological ranges. This suggests that the wave propagation induced by layer sliding does not only apply to our specific experimental setting but also work for a broader range of biological systems on various scales.

\section*{Results}
\subsection*{Concentric wave emergent with weakly adhesive substrate layers}
We cultured an epithelial monolayer on a gel plated on a glass base (Fig. \ref{fig:scheme}A). By weakening the adhesion at the gel-glass interface, we can allow the gel to slide on the glass and hence the deformation of the gel substrate will depend on time. To introduce a force event in the cell layer, we applied a UV laser to damage the DNA of a single cell at the center and to induce its apoptosis. At a few minutes after ablation, a concentric wave of deformation was observed in the substrate (Fig. \ref{fig:scheme}B, top and middle; Movie S1). 

The deformation dynamics of the gel was captured by the fluorescent beads embedded at the surface using traction force microscopy (TFM). We tracked fluorescent beads on the images (Fig. \ref{fig:scheme}B, top) and measured a radial displacement for each bead (Fig. \ref{fig:scheme}B, middle), with the origin being defined by the apoptotic cell. By averaging the radial displacement over the beads located in the distance $\sim r$ from the center, we obtained the scalar displacement field $u_r(r)$ (red curves in the bottom panel of Fig. \ref{fig:scheme}B; See Materials and Methods for beads tracking and measurement; See Movie S2 for the dynamics of beads). 

The concentric wave was characterized by a strong peak in the bead displacement field $u_r(r)$ moving outward from the cell extrusion point. The wave moved slowly over a distance of 2 cell diameters in the first 10 minutes (Fig. \ref{fig:scheme}B, bottom). 
This happened even with a very thick substrate, {\it e.g.} with $\sim 50 \ \mu$m thickness, which is larger than a typical cell diameter. 
The peak separates the space into two regions with inward and outward movement of beads (Fig. \ref{fig:scheme}C) and 
nearly 40 mins after ablation, the bead displacement field starts to stabilize. 

By contrast, if the gel was strongly adhesive to the glass so that there is no interfacial sliding between two layers (which we call hereafter ``non-slidable" case), then the radial displacement of the beads exhibited no peak or propagation dynamics (Fig. \ref{fig:scheme}D) and the magnitude of the displacement is about 10 times smaller than that observed in a slidable case (Fig. \ref{fig:scheme}B, bottom).

\subsection*{Diffusive dynamics of peak propagation}
The remarkable feature of bead displacement $u_r(r)$ is the emergence of a strong peak.
Fig. \ref{fig:peak}A(insets) shows the time evolution of the position of peak $r_{peak}$ and the peak value $u_{peak}$ for various samples. After rescaling (see Materials and methods),
$r_{peak}$ and $u_{peak}$ collapse to $\tilde{r}_{peak}$ and $\tilde{u}_{peak}$, where two stages of propagation could be found (Fig. \ref{fig:peak}A). The first stage (stage I, $< 20$ min) is characterized by a diffusive increase in both $r_{peak}$ and $u_{peak}$ and the second stage (stage II, $>20$ min) by the saturation of $r_{peak}$ and the decrease in $u_{peak}$. 

During stage I, there exists a negative correlation between the magnitude of peak and the propagation speed characterized by $v_{peak}(t)=\Delta r_{peak}(t)/\Delta t$ ($\Delta$ denotes the difference between two consecutive frames along time). As shown in Fig. \ref{fig:peak}B, $v_{peak}(t)$ calculated over all the samples in stage I is plotted against the magnitude of peak $u_{peak}(t)$, and the black straight line corresponds to $v_{peak}(t)u_{peak}(t)=0.4 \mu \mathrm{m}^2/\mathrm{min}$. This infers that a faster propagation of the deformation corresponds to a smaller extent of the deformation. 

Another feature of the displacement field worth noticing is the slopes of the rise and decay around the peak. Figure~\ref{fig:peak}C exhibits a typical sample plot of its smoothed displacement over 11 minutes after apoptosis in a log-log plot. The near-center displacement $u_r(r<r_{peak})$ grows essentially in a linear way with respect to $r$, whereas the slope of decay in the displacement field $u_r(r>r_{peak})$ follows essentially a $1/r$ dependency. To investigate the tails exponent of $u_r$ quantitatively, we plot the rescaled displacement $\tilde{u}_r=u_r/u_{peak}$ for multiple samples in Fig. \ref{fig:peak}D. Experimental noises from optical defects (which we estimated to be on the order of 0.03 $\mu$m) have been removed from the data and thus improving the quality of the fit. The inset of Fig. \ref{fig:peak}D shows the sample-wise fitted tail powers which has a mean $-1.00\pm 0.18 $ (90\% confidence interval, two gray lines).

\subsection*{3D elastic gel model}
%Next, we address the physical mechanism generating this long-range concentric propagation of gel deformation field based on linear elasticity theory.  
Referring to the experimental results, we construct our theory focusing on the gel mechanics and consider the forces exerted by the tissue as a boundary condition. %This decoupling of gel mechanics from tissue mechanics facilitates an investigation of the roles of ECM and tissue separately in the wave propagation.
The creeping time of the PDMS gel used in our experiments is roughly 2 seconds \cite{kawaue,leong2015viscoelastic}, which is far shorter than our timescale of interest. Therefore, we model the gel as a pure elastic material. Considering the axisymmetry in our problem, the model is set in cylindrical coordinates (Fig. \ref{fig:scheme}E). Since the beads in experiments were observed in terms of radial displacements, here, we only explain the radial deformation part in the model  (see details of model construction in section S1 ). 

We construct the theory relying on the linear elasticity of an incompressible gel. The radial displacement field $u_r(r,z,t)$ is a function of radial distance from the center $r$, the vertical position $z<h$, and the time $t$, where $h$ is the height of the gel.
According to the force balances inside the gel and the incompressibility condition, the spatial profile of $u_r$ obeys
\begin{equation}
    \frac{\partial^2 u_r}{\partial r^2} + \frac{1}{r}\frac{\partial u_r}{\partial r} - \frac{u_r}{r^2} + \frac{\partial^2 u_r}{\partial z^2} 
    - 
    \frac{2}{\hat{G}} \frac{\partial P}{\partial r} = 0 ,
\label{eq:ur_3d}
\end{equation}
where $P$ is the pressure field, which is given through the incompressibility condition, and $\hat{G}$ is an effective shear modulus of the gel.
 
The only external forces are the shearing forces
on the surface $z=h$ and at the bottom $z=0$:
\begin{equation}
    \left\{
    \begin{aligned}
        \sigma_{rz}\big|_{z=h}&=S(r,t)\\
        \sigma_{rz}\big|_{z=0}&=\xi\frac{\partial u_r}{\partial t}\bigg|_{z=0}\\
    \end{aligned}
    \right.
    \label{eq:boundary} , 
\end{equation}
where $\sigma_{rz}$ is the shear stress defined as
\begin{equation}
    \sigma_{rz}= \frac{\hat{G}}{2} \left( \frac{\partial u_r}{\partial z} + \frac{\partial u_z}{\partial r}\right) \ 
    \label{sigma_rz}
\end{equation}
with the vertical displacement $u_z(r,z,t)$. 
On the gel surface, $S(r,t)$ is an unknown form of stress field exerted by the cells and we name this term $S(r,t)$ ``surface stress" in what follows. At the bottom of the gel, $\xi$ is the friction coefficient that describes the force-velocity relationship during the gel sliding on the glass.
For the $z=h$ condition, we have assumed that the change of height over time and space is negligible compared to the magnitude of $h$ itself and thus ignored the tilt of the surface. We also fix $h$ to be a constant.

With specific boundary conditions in our problem, 
%we could the general solution form
the general solution is derived
 (derivations in section S1) as
\begin{equation}
    u_r(r,z,t)=\frac{1}{2\pi}S(r,t)***M(r,z,t),
\end{equation}
where the operator ``***" denotes a space(2D)-time convolution of the surface stress $S(r,t)$ and a memory kernel $M(r,z,t)$. 
%The memory kernel describes the impulse response of the gel deformation with the bottom friction, which is irrelevant of the specific form of the surface stress. and 
The kernel $M(r,z,t)$ describes the response in a gel deformation against the ring-shaped impulse in radial directions at the top surface of the gel and is independent of the specific form of $S(r,t)$. Even though the gel is assumed to be elastic, the dependence on time $t$ is brought into the kernel by the friction at the bottom of the gel (Eq.\ref{eq:boundary}).
The specific form of $M(r,z,t)$ is shown in Materials and Methods and section S1.1.
A detailed analysis of the memory kernel (section S3.1) reveals that if the elapsed time $t$ surpasses a critical timescale $t_c= h\xi/\hat{G}$, the $z$-dependency in the solution becomes negligible, so the solution is approaching to a 2D limit form $u_{r,2D}(r,t)=S(r,t)***M_{2D}(r,t)$, and $M_{2D}$ could exhibit a dynamic scaling, which is the origin of the propagation dynamics in the substrate. 

In contrast, if the elapsed time is significantly shorter than the critical timescale $t_c$, i.e., $t\ll h\xi/\hat{G}$, the final solution can be reduced to a specific form $u_r^n(r,z,t)$, which is just the immediate deformation under the shearing source $S(r,t)$. Note that the short time limit $t\ll h\xi/\hat{G}$ is also equivalent to a large friction limit $\xi\gg \hat{G}t/h$, or the thick gel limit $h\ll t\hat{G}/\xi$. Therefore, $u_r^n$ is also the solution under a non-slidable setting of gel (rigidly bonded to the glass bases). In this limit, there is no transmission of forces at long distances and all the dynamics in the substrate is instantaneous and local as expressed by $S(r,t)$.

\subsection*{Models of cell mechanics}

The external stress field at the surface $S(r,t)$ is transmitted from the tissue through focal adhesion patches. Choosing a form for this term implies a corresponding hypothesis for tissue biomechanics. Evidence shows that such force is not provided by the crawling of cells \cite{kawaue}. The breakdown of intercellular junctions could be a viable mechanism to release the stress previously stored in those junctions on a very short time-scale and thus push the substances away from the apoptotic cell \cite{Teng2017}. This transient process generates a spatial stress distribution which transmits to the gel surface. The same experiment also found that the sliding of focal adhesions is negligible at the interface between the cell layer and the gel, indicating that the force is transmitted through the elastic deformation of integrins and hence the external stress could persist for some time\cite{kawaue}. Since it is unknown how the stress is distributed in space and evolves over time due to complex biochemical processes upon an apoptosis event, we start from a simple and reasonable $S(r,t)$ as a benchmark, and see to what extent the experimental observations can be recapitulated. 

We first choose a power-law dependence for the decay of $S(r,t)$ over the distance $r$ from the center:
\begin{equation}
    S(r,t)=s \left(\frac{\varepsilon}{r}\right)^q \Theta(r-r_0) \Theta{(t-t_0)},
\label{eq:spreading}
\end{equation}
where $s$ is the coefficient for the magnitude of prestress in tissue, $r_0$ is the onset position of this stress, $t_0$ is the time for the release of stress, and $\varepsilon$ is the cell radius, which also determines the scale of spreading distance of the force. The function  $\Theta$ is a Heaviside step function. Since the roles of $t_0$ is trivial in the calculation, we set $t_0=0$ in the following demonstration. The role of $r_0$ is case-dependent, and we leave the explanations in S3.4. Here we assume the simplest case where $r_0=0$. The reason why we use power law decay instead of others such as exponential ones is to avoid additional length scale at this moment. The power $q$ is chosen to be $0<q<3$ to guarantee the existence of the solution. 

%Specific solution

Figure \ref{fig:theory}A shows the numerical results of displacement fields near the surface $u_r(r,z=0.9h,t)$ (red) for the case of $q=1$ with dimensionless time $\tilde{t}=t/t_c=t\hat{G}/h\xi$. For comparison, the solution under non-slidable conditions $u_r^n(r,z=0.9h,t=0)$ and under 2D approximation $u_{r,2D}(r,t)$ are also shown in blue dashed curves and black curves. For $t<t_c$, the displacement field $u_r(z=0.9h)$ (red curves) has a peak magnitude close to  $s\varepsilon/\hat{G}$ in the near field $r<h$, the same as that with a non-slidable condition (blue curves) and this magnitude will increase with $z$ (for a better intuition on the $z$ dependence of the solution, see Fig. \ref{fig:theorySimple}B in supplementary materials). Yet, this displacement is transient and caused by the shear deformation of the gel. When $t$ surpasses $t_c$, the far field of $u_r(r>h, z=0.9h)$ transitions dynamically from $u_r^n(r>h)$ to $u_{r,2D}(r>h)$. Notice that all the tails of these solutions decay as $1/r$, which agrees well with the tail power $-1$ found in experiments (Fig. \ref{fig:peak}, C and D). 

The bottom panel of Fig. \ref{fig:theory}A shows that the peak positions and heights of $u_r(z=0.9h)$ (red marks and curve) and $u_{r,2D}$ (black marks and curves) collapse to the same curves which grow diffusively with time when $t\gg t_c$ and a propagating peak emerges from $r\sim h$. These results suggest that a 2D approximation limit ($t>t_c$) is sufficient to explain the  peak-propagation dynamics at the late time regime. 

Therefore, for the sake of simplicity and for our purpose of studying the peak propagation phenomena, we would hereafter focus on the dynamics of $u_{r,2D}$ which is amenable to analytic calculations. The analytical form of $u_{r,2D}$ is the convolution of the source $S(r)$ with the memory kernel $M_{2D}$ (calculations in section S2.1):
\begin{equation}
    \label{eq:solution2d}
        u_{r,2D}(r,t)\sim
        \frac{s\varepsilon^q}{2\hat{G} h}r^{2-q}\Theta \left(\sqrt{\frac{2\hat{G} ht}{\xi}}-r\right)+\frac{s\varepsilon^q t}{2\xi r^q}\Theta \left(r-\sqrt{\frac{2\hat{G} ht}{\xi}}\right).
\end{equation}

A propagation of the peak is found with $0 < q < 2$. Particularly, $u_{r,2D}$ grows with $r^{2-q}$ for $r<r_{peak}$ and decays with $1/r^q$ for $r>r_{peak}$ in space and a crossover from the growth to decay occurs near $r\sim r_{peak}$. As seen in Eq.\ref{eq:solution2d}, the power of the decaying tail depends on the choice of $q$. Among them, for $q=1$, the tail power is $-1$, which indeed matches the result shown in the top panel of Fig. \ref{fig:theory}A.
Therefore, through the comparison with the experimental observation (Fig. \ref{fig:peak}D), we fix $q$ to be $1$ in the following.

The peak of displacement $u_r(r)$ locates at
\begin{equation}
    r_{peak}(t)= \sqrt{\frac{ 2\hat{G} ht}{\xi}},
\end{equation}
 and if $q=1$, the magnitude of peak is (calculations in section S3.2)
\begin{equation}
     u_{peak}(t)\sim Z s\varepsilon \sqrt{\frac{t}{2 \hat{G} h\xi}},
     \label{eq:u_peak}
\end{equation}
where
\begin{equation}
    Z=\int_0^\infty dx J_1(x)\frac{1-e^{-x^2}}{x^2}\sim 0.48227,
\end{equation}
with $J_1$ the Bessel function of the first kind of order 1. Both the dynamics of peak position and magnitude 
grow with $\sqrt{t}$. 
Accordingly, the speed of peak propagation is
\begin{equation}
    v_{peak}=\sqrt{\frac{\hat{G} h}{2 t\xi}},
\end{equation}
which is enhanced by a slipperiness of gel base (smaller $\xi$) or a large bulk stiffness of gel (larger $\hat{G}h$) and meanwhile, the speed of peak propagation slowdowns with time. 

Moreover, the speed is inversely proportional to $u_{peak}$ independently of time as
\begin{equation}
\label{eq:negativeCorrelation}
   v_{peak} = \frac{Zs\varepsilon}{2\xi}
   u_{peak}^{-1}.
\end{equation}
This reciprocal relationship between speed and magnitude of the peak implies that a faster propagation corresponds to a smaller peak, which agrees with the experimental facts well (Fig. \ref{fig:peak}B).

Interested readers can refer to Figure \ref{fig:s-dep}B for a comparison of the propagation dynamics under several simple forms of $S(r)$. The nature of propagation dynamics is well preserved under other forms of $S(r)$ but the propagating profiles are distinct from the peak profiles found with $S(r)\sim 1/r^q$.

In our experiments, we also observed the slowdown of propagation and decay in the peak magnitude at the second stage ($>20$ min). This deviation at such late stage from the simple surface stress results could also be explained by our theory if the spatial distribution and dynamics of the surface stress field $S(r,t)$ is comprehensively modeled (see sections S2.2 and S2.3, in which we provide its possible examples).

\subsection*{Quantitative validation}
Here, we investigate on a quantitative level whether our theory with the simple surface stress profile (Eq.\ref{eq:spreading}) can explain the experimental observations for the early time regime ($t<15$min) based on the following experimental measurable parameters 
\begin{equation}
    \varepsilon = 10 \ \mu\mathrm{m},\ \hat{G} = 10 \ \mathrm{kPa}, \  h = 50 \ \mu \mathrm{m}.
    \label{parameters}  
\end{equation}
The first crucial test for the model validity is to predict the case where the friction is too large for the gel to slide.
Under a simple surface stress field (Eq.\ref{eq:spreading}, $q=1$), the displacement field under non-slidable condition is a steady distribution and with its peak magnitude maximized at the surface:
\begin{equation}
    u_r^n(r,z=h) \sim \frac{s\varepsilon}{\hat{G}} 
    \label{eq:solNonSlipSpreading}
\end{equation}
(calculations in section S3.3), which is constrained solely by the ratio between cell's contractility $s$ and gel's elasticity $\hat{G}$. 

Previous literature shows that the magnitude of tissue prestress $s$ ranges from $10^2\sim 10^4$ Pa \cite{Tambe2011,wang2002cell}. Given the aforementioned parameter values (Eq.\ref{parameters}), the magnitude of $u^n_r$ should result in a range as:
\begin{equation}
     0.1\mu \mathrm{m}<u^n_r < 10 \mu\mathrm{m}.
     \label{eq:un_r}
\end{equation}
Experimentally, the measured magnitude of bead displacement is also constrained by the available microscope resolution. If $u^n_r(r)$ is below the resolution threshold, the information about the motions of some beads would be entangled with optical noises that would undermine the quality of the measurement. Figure \ref{fig:disc}A shows typical snapshots of bead displacement with a strongly adhesive (non-slidable) gel in experiments, where the magnitude of averaged bead displacement (red) is far smaller than the resolution (0.206 $\mu$m/px) with a large standard deviation (See Movie S3 also). This indicates that the upper bound of the magnitude of $u^n_r$ should be $\sim0.1\mu$m. Combining Eq.\ref{eq:un_r}, we conjecture $u^n_r\sim 0.1\mu$m with the tissue prestress $s\sim 10^2$ Pa in our experiments.

Next, we test the validity our model in the weakly adhesive case. Based on Eqs.\ref{eq:solNonSlipSpreading} and \ref{eq:u_peak}, the magnitude of the ratio between the peak value with a
weakly adhesive gel $u_{peak}(t)$
and the value with
a rigid gel $u^n_r$ is approximately 
\begin{equation}
\label{eq:MagnitudeDifference}
    \frac{u_{peak}(t)}{u^n_r}\sim \frac{1}{2}\sqrt{\frac{\hat{G} t}{h\xi}},
\end{equation}
which is about 10, as shown in Fig. \ref{fig:disc}B.
By applying Eq.\ref{eq:MagnitudeDifference}, we could also estimate the unknown friction coefficient between gel and glass $\xi$ for the weakly-adhesive case as
\begin{equation}
\xi \sim 2\times 10^{10} \ \mathrm{Pa}\cdot \mathrm{s/m},
\label{eq:non-slidable_eval}
\end{equation}
which agrees with the magnitudes estimated in previous literature \cite{kruse2006contractility}.
Then, from Eq.\ref{eq:negativeCorrelation} with the parameter values given in Eqs.\ref{parameters} and \ref{eq:non-slidable_eval}, the multiplication of the propagation speed $v_{peak}$ and the magnitude of peak $u_{peak}$ is a constant independent of time estimated as
\begin{equation}
    v_{peak} u_{peak} \sim \frac{s\varepsilon}{4\xi}
    \sim  10^{-14} \ \mathrm{m^2/s},
    \label{eq:reciprocal}    
\end{equation} 
which agrees with the magnitude $10^{-14}\mathrm{m^2/s}$ found with stage I in experiments (Fig. \ref{fig:peak}B). 
Furthermore, the critical timescale $t_c$ for the onset of propagation is 
\begin{equation}
    t_c\sim\frac{\xi h}{\hat{G}} = 50 \ \mathrm{s}\sim 1 \ \mathrm{min},
    \label{eq:onsetTime}
\end{equation}
which means this onset of propagation is critically invisible in our experiment with time resolutions on a minute-scale. All these quantitative agreements with the experiments confirm the validity of our model.

\subsection*{Diversified scale of propagation and deformation}
\begin{figure}
    \centering
    \includegraphics[width=\linewidth]{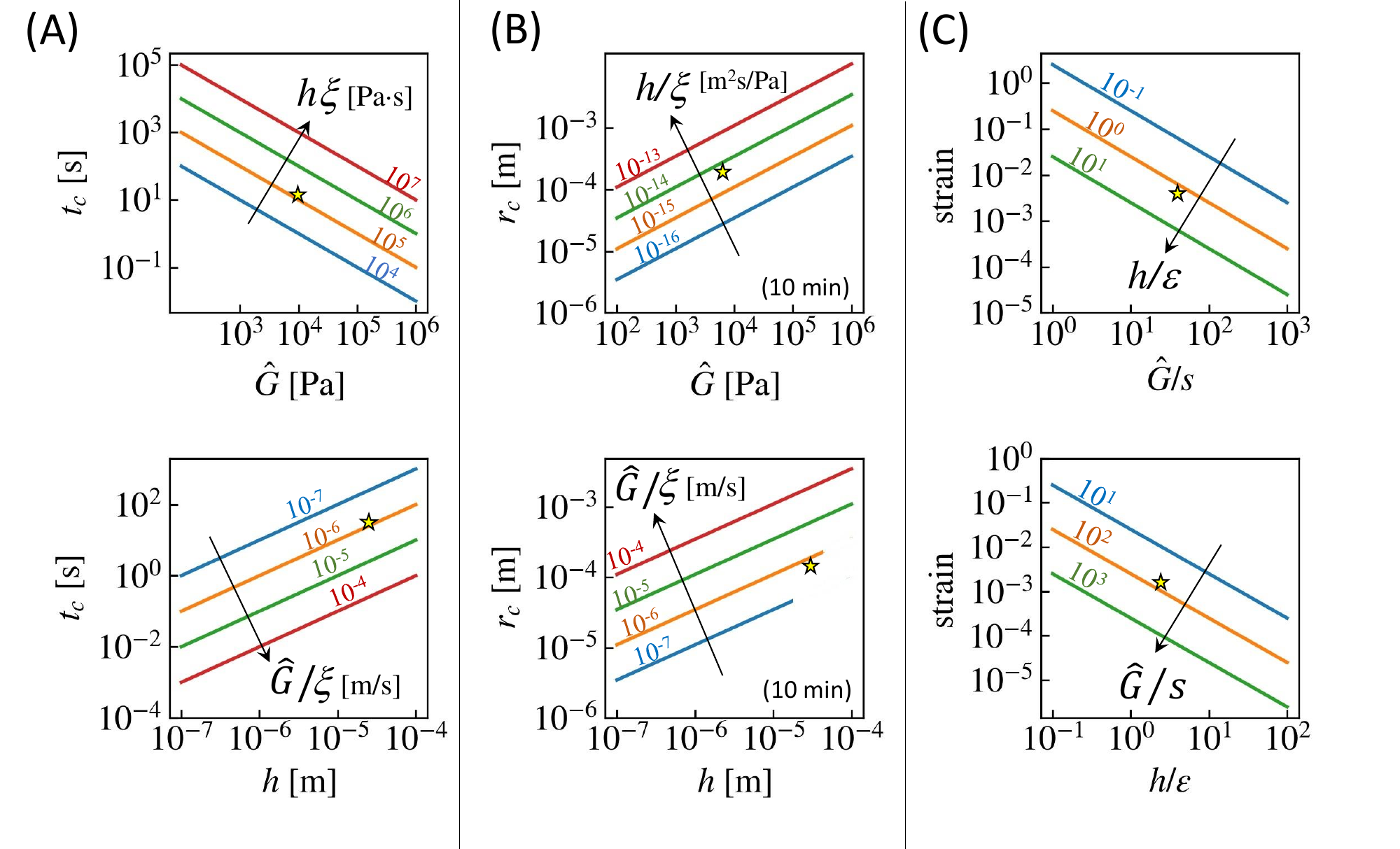}
    \caption{Diversified scale of propagation. (A) The observation timescale $t_c=\xi h/\hat{G}$. (B) Magnitude of propagation distance $r_{c}$ at 10 minutes (for $t_c\ll $10 min). Note that $r_c=r_{peak}$ under the persistent decay form of the surface stress $S(r)$. (C) Magnitude of strain calculated by $u_{peak}/r_{peak}$  (angular strain at peak, which is independent of time $t$) under the persistent surface stress $S(r,t)=s\varepsilon/r \Theta(t)$. The stars indicate the parameter settings in our experiments, which result in $t_c \sim 1$min, $r_c \sim 100 \mu$m, strain $\sim 10^{-2}$.  }
    \label{fig:parameter}
\end{figure}

Although the propagation found in our experiments manifests on a scale of 50 $\mu$m/10 min, our theory predicts that this propagation scale could occur on multiple scales if the mechanical properties of the substrate layer are changed. Fig. \ref{fig:parameter} shows how the observation timescale $t_c$, propagation distance $r_{c}$ and strain of deformation varies with the stiffness $\hat{G}$, thickness $h$ and friction ($\xi$) of the gel tuned to be within physiological ranges. The timescale $t_c$ is the proper time resolution for observing this propagation dynamics. This timescale ranges from less than 1 second to several hours (Fig. \ref{fig:parameter}A). The propagation distance $r_c=\sqrt{2\hat{G}ht/\xi}$ (see Eq.\ref{eq:Mf_scaling}) at $t=$10 min ranges from micron to millimeter for $t_c<10$min (Fig. \ref{fig:parameter}B). 

The extent to which the substrate surface is deformed is biologically relevant because cells can sense the local strain of the substrate and trigger mechanotransduction processes accordingly. The magnitude of strain not only depends on the mechanical properties of the substrate but also on the prestress magnitude $s$ and cells size $\varepsilon$ of the tissues. Under persistent decaying surface stress $S(r,t)=s\varepsilon/r (t>0)$, the strain, which is calculated by the ratio between peak magnitude and peak position $u_{peak}/r_{peak}\sim s \varepsilon/4\hat{G}h$, also varies drastically across from $10^{-1}$ to $10^{-5}$ to with the increase in the normalized the layer stiffness $\hat{G}/s$ and with the increase of the normalized layer thickness $h/\varepsilon$ (Fig. \ref{fig:parameter}C).

\subsection*{Physics of force propagation through interfacial sliding} 
We next analyze our theory in more pedagogical way.
The essence of propagation lies in the diffusive nature of $u_r$ dynamics under a non-vanishing spreading source (see Eq.\ref{solslip} in section S1.1):
$
   \partial u_r/\partial t = \mathrm{source}(r,t) - \mathrm{coff.} \times \Delta u_r
$,
%\end{equation}
in our model and the diffusive dynamics comes from the force balance between gel's elasticity and frictional sliding (see an analysis of a simplified model without the pressure term in section S4).

%\begin{equation}

If we inspect the evolution of radial displacement $u_r(r)$ at a local position $r$ with a weakly-adhesive base under a persisting surface stress field $S(r)$ (Fig. \ref{fig:disc}C, top), we find that $u_r(r)$ at a given $r$ has a sigmoid-like growth as it starts from an initial non-zero value and gradually saturates at another value over time. The initial regimes should correspond to the shear deformation with a fixed bottom boundary and the final regimes should correspond to the bulk stretch for which the gel's sliding at the bottom is completely relaxed. The separation of stretching and shearing regimes depends on time and space. Where $r \ll \sqrt{2\hat{G}/(\xi h)} \times t^{1/2}$, stretching becomes dominant (See Eq.\ref{eq:M_2D} in section S2). This suggests that the regime of $u_{\mathrm{stretch}}$ expands out from $r=0$ like normal diffusion (Fig. \ref{fig:disc}D top and middle). The emergence of the peak is a corollary of this regime separation in space. As long as we assume a decaying form of Eq.\ref{eq:spreading} with $0<q<2$, the displacement $u_{\mathrm{stretch}}(r)$ in stretching regime has an opposite trend in $r$-dependence against that in shearing regime $u_{\mathrm{shear}}$ (see calculations in S2.1). Consequently a strong peak appears at the transition boundary from the stretching to shearing regimes in space (Fig. \ref{fig:disc}D bottom).

Finally, if regarding $u_r(r,h)/h$ in Fig. \ref{fig:disc}C as an effective shear strain in the gel, we can find its time-evolution at any local position $r$ behaves like the creeping of a rigidly bound substrate with viscoelasticity in a motif named ``Zener model" (Fig. \ref{fig:disc}C, bottom). The creeping time is proportional to $r^2\xi/\hat{G}h$. This similarity in the creeping dynamics indicates that our system which features a structure that allows interfacial sliding between layers can be analogous to a viscoelastic substrate able to creep under applied forces, but at a much larger scale than the natural creeping timescale of the materials. For instance, the gel in our experiments has a creeping timescale of roughly 2 seconds \cite{kawaue,leong2015viscoelastic} but the wave propagation dynamics last more than 10 minutes.

\section*{Discussion}
\subsection*{Summary} 
In this article, we studied how the layered structures that allows interfacial sliding in a substrate can help propagate mechanical cues and reported on a novel form of long-range force propagation in epithelial tissues through its substrate \textit{in vitro }. This was observed experimentally in settings composed of a single epithelial layer adhering strongly to an elastic gel which itself was adhering weakly to a rigid substrate. A concentric wave was generated in the elastic gel by the ablation of a single cell in the tissue, and the subsequent propagation lasted tens of minutes across several cell diameters (Fig. \ref{fig:scheme}A to C, and Fig. \ref{fig:peak}).  This observation can be quantitatively understood by studying theoretically the dynamics of an elastic gel adhering strongly to the tissue but which is able to slide on the rigid substrate (Fig. \ref{fig:scheme}D and Fig. \ref{fig:theory}). The ablation process generates a force field essentially shearing the elastic gel and the propagation of this deformation is allowed by the sliding process on the rigid substrate side (Fig. \ref{fig:disc}). The theory further revealed that the spatiotemporal scale on which the propagation occurs can vary across multiple orders of magnitude if we tune the stiffness, thickness and slipperiness of the substrate layers within physiological ranges, whereas the magnitude of local strain of the substrate depends on the tissue prestress and cell size as well (Fig. \ref{fig:parameter}). This variability of the propagation scales suggests that this type of force propagation can function at different biological levels.

Different from those mechanical oscillations realized through viscoelastic properties of tissues \cite{serra2012mechanical,deforet2014emergence,blanch2017hydrodynamic}, the concentric wave in our experiments propagates through the elastic substrate, even with a steady stress field in tissue. The wave is peculiarly characterized by a pronounced peak, whose magnitude inversely proportional to the propagation speed.

\subsection*{Dynamic remodeling of substrate as an origin of layer friction}
%Finally we should stress the importance of dynamic remodeling of the substrate structure. 
A weakly adhesive substrate that facilitates interfacial layer sliding is central to the emergence of this novel form of force propagation. % \textit{in vitro}.
Here we discuss the reason why this key experimental setting and theoretical assumption in this work could be relevant in a real living organism. 
\textit{In vivo}, the tissue substrates are composed of multiple layers of ECM \cite{lu2012extracellular} such as basement membrane and lamina propria and other connective tissues. The biopolymer linkages between these layers would not be as rigid as those prepared in conventional TFM experiments because of continuous ECM remodeling. The turnover of attachment-detachment dynamics of linkage components will inevitably cause ``frictional sliding" at layer boundaries when external stresses are applied: the more frequent the turnovers are, the smaller the friction.

However, to qualitatively connect the experimental and theoretical results described in this paper to the \textit{in-vivo} situation will require to further knowledge. We have to calibrate the mechanical parameters like interlayer friction for the tissues in actual living organisms. Analyzing the turnover rates of attachment-detachment of linkers based on molecular chemistry is possible method to evaluate the layer friction \cite{TAWADA1991193}. % for simple structures(???) and apply our model.} However,
Realistic ECM remodeling will depend on the system constituents 
and geometry, and may exhibit diverse kinds of inhomogeneity and anisotropy. How we can adapt the frictional dynamics in the current model to more generic types of ECM remodeling is a theoretical challenge for future investigation.

\subsection*{Cell responses to substrate sliding} 
As shown previously from epithelial-monolayer experiments, the frictional sliding at the gel-glass interfaces promotes the Yap nuclear translocation rate and cell proliferation around the extruding cell~\cite{kawaue}. 
%We can also explain the phenomena with our theory here.
Since cells may be able to sense the substrate strain instead of the displacement field itself, here we add comments on strain field dynamics associated with force propagation induced by substrate layer sliding. 
%It has been understood 
Indeed, there is literature reporting
that local substrate strain dictates the cell-substrate mechanosensing~\cite{panzetta2019cell}. 
%In our problem, the the force propagation induced by substrate layer sliding could be sensed by the cells through the strain field dynamics at the cell-substrate interface, which are :
In our theory, the strain field is given by
\begin{equation}
    \mathrm{radial \ strain:\ }\epsilon_{rr}=\frac{\partial u_r}{\partial r},\ 
    \mathrm{angular\ strain:\ }\epsilon_{\varphi\varphi}=\frac{u_r}{r}.
\end{equation}
Near the center ($r<r_{peak}$), cells can sense a local stretching strain in both the radial and angular directions ($\epsilon_{rr} > 0,\epsilon_{\varphi\varphi}>0$), whereas far from center ($r>r_{peak}$), cells sense the expansion of the substrate along the angular direction while they sense the contraction of the substrate in the radial direction ($\epsilon_{rr}<0$ and $\epsilon_{\varphi\varphi}>0$, see Fig. \ref{fig:strain}). Since cells are especially sensitive to local stretching strain in the substrate for regulating their behaviors \cite{panzetta2019cell}, mechanosensing activities such as YAP nuclear translocation \cite{dupont2011role,nardone2017yap} are expected within this stretched substrate region. 

The cellular responses to the substrate deformation can reciprocally modify major contributors to the force propagation through substrate such as the cutoff range, the temporal evolution, or the spatial distribution of stresses in tissues. Further theoretical and experimental investigations on the effect of mutual influence between the cell's responses and substrate dynamics can improve our understanding on the biological significance of interfacial friction among tissue layers \textit{in vivo}.

\section*{Materials and Methods}
\subsection*{Cell extrusion experiment}
A monolayer of MDCK cells or MKN28 cells was cultured on a PDMS gel (CY52-276A:CY52-276B=1:1, Dow Corning) overlaid on a glass-bottom Petri dishes (IWAKI). The thickness of the gel is 50 $\mu$m, and its Young’s Modulus is 15kPa with a Poisson ratio 0.499. Fluorescent (red or far-red, Invitrogen) beads are embedded at the surface of the gel to capture the displacement of the gel surface. A strongly adhesive (non-slidable) base of a gel was prepared by silanizing a glass-bottom Petri dishes with 3-aminopropyl trimethoxysilane (APTES, Sigma). A weakly adhesive (slidable) base of a gel was prepared without APTES. Cells and beads were imaged with a NikonA1R MP laser scanning confocal microscope with Nikon Apo 60x/1.40 oil-immersion objective. Apoptosis was induced using a UV laser as described before \cite{KOCGOZLU20162942}. The summary of the experimental setting is summarized in Table 1. Full details of the experimental protocol can be found in \cite{kawaue}.

%%%
\subsection*{Calculation of radial displacement}
We first preprocessed the images of fluorescent beads for a proper contrast and uniform intensity in grayscale to suppress the impact from optical defects as much as possible. Then, we tracked each spot in the images by a Python package \texttt{trackpy}, and calculate the vector field of displacement from time=0, as shown in Fig. \ref{fig:scheme}B. Then we projected all the vectors $\vec{u}(\vec{x})$ at different position $\vec{x}=(x,y)$ onto its radial direction to get the radial displacement $u_{r}= \vec{u}\cdot \vec{x}/|\vec{x}|$ and turn the vector field to a scalar field $u_{r}(\vec{x})$. Considering the symmetry about the center (where the apoptotic cell lies), this 2D scalar field can be further reduced to a 1D field $u_r(r)$ by averaging the bead displacement for beads located at a distance from center $r$ with a range $\Delta r\sim 10\mu$m. Finally, we perform a Savitzky-Golay filter with subset size $\sim 25\mu$m and polynomial order 3 to smooth the averaged displacement.

\subsection*{Rescaling of peak dynamics}
The propagation slows down with time t and the evolution of peak position $r_{peak}$ can be well fitted by a function of the following form
\begin{equation}
    r_{peak}(t)=r_{peak}^{\infty}\big(1-e^{-\sqrt{t-t_0}/\tau}\big)
    \label{eq:fitting}
\end{equation}
where $r_{peak}^{\infty}$ is the furthest position the peak can reach, $t$ is the onset time of the deformation and $\tau$ is the timescale for the peak to stop propagation. These three parameters are found through fitting the data for each sample. 

The peak positions $r_{peak}(t)$ of all samples then collapse to a rescaled peak position $\tilde{r}_{peak}$ with these fitting parameters (Fig. \ref{fig:peak}A, top):
\begin{equation}
    \tilde{r}_{peak}(t)=\frac{r_{peak}\left(\frac{t-t_0}{\tau}\right)}{r^\infty_{peak}}\sim 1-e^{-\sqrt{(t-t_0)/\tau}}.
\label{scaling_r}
\end{equation}

Similarly, the corresponding peak height $u_{peak}$ can be rescaled using $t_0$ and $\tau$ fitted from $r_{peak}$ as
\begin{equation}
\tilde{u}_{peak}(r,t)=\frac{u_{peak}\left(\frac{t-t_0}{\tau}\right)}{u^\infty_{peak}}
\label{scaling_u}
\end{equation}
where $u^\infty_{peak}$ is the fitted maximal value for $u_{peak}$. As shown in the bottom panel of Fig. \ref{fig:peak}A, the rescaled peak heights $\tilde{u}_{peak}$ collapse to a nonmonontonic trend: for time $t-t_0\ll \tau$,  $\tilde{u}_{peak}$ grows in a diffusive way (dashed line), i.e., $\propto \sqrt{t}$; yet, for time $t-t_0\gg \tau$, the $\tilde{u}_{peak}$ slowly decreases. Therefore, $\tau$ separates the whole process into two stages in time. 

%%%%
\subsection*{Theory}
The essences of the theoretical model used in this paper is given in the main text. The memory kernel $M(r,z,t)$ derived from the model is given as the inverse Hankel transform of order 1 (from the wave-number space $k$ to the real-space $r$) of the following form of function: 
\begin{equation*}
M(k,z,t) =
m_s(k,z) \delta (t) + m_f(k,z) M_0(k,t) \ .
\end{equation*}
The first term stems from purely the gel's elastic property. This term has a time-dependency represented by a Dirac's delta function $\delta(t)$, and the coefficient $m_s(k,z)$ is independent of friction. 
The second term incorporates the effect of friction at the bottom of the gel. This term has a finite-range time dependency on $M_0(k,t)$, and the coefficient $m_f(k,z)$ depends on friction.
In this way, one can analyze the contribution of friction in the memory kernel, which enables us to investigate the detailed mechanism behind the gel dynamics. The explicit functional forms of $m_s(k,z)$, $M_0(k,t)$ and $m_f(k,z)$ and the full model and analysis are elaborated in Supplementary material (section S1). 

\subsection*{Statistical analyses}
The fitting procedures with the statistical analyses in Fig. \ref{fig:peak} were carried out by a Python package \texttt{SciPy}. The error bars in Fig.\ref{fig:scheme}B and Fig.\ref{fig:disc}A were calculated as standard deviations of the data points within a moving window length about $15\mu$m.

%BibTeX users: After compilation, comment out the following two lines and paste in
% the generated .bbl file. 

\bibliography{citationsIntroduction}

\begin{thebibliography}{10}
\expandafter\ifx\csname url\endcsname\relax
  \def\url#1{\texttt{#1}}\fi
\expandafter\ifx\csname urlprefix\endcsname\relax\def\urlprefix{URL }\fi
\providecommand{\bibinfo}[2]{#2}
\providecommand{\eprint}[2][]{\url{#2}}

\bibitem{De2007}
\bibinfo{author}{De, R.}, \bibinfo{author}{Zemel, A.} \&
  \bibinfo{author}{Safran, S.~A.}
\newblock \bibinfo{title}{Dynamics of cell orientation}.
\newblock \emph{\bibinfo{journal}{Nature Physics}}
  \textbf{\bibinfo{volume}{3}}, \bibinfo{pages}{655--659}
  (\bibinfo{year}{2007}).
\newblock \urlprefix\url{https://doi.org/10.1038/nphys680}.

\bibitem{PhysRevE.78.031923}
\bibinfo{author}{De, R.} \& \bibinfo{author}{Safran, S.~A.}
\newblock \bibinfo{title}{Dynamical theory of active cellular response to
  external stress}.
\newblock \emph{\bibinfo{journal}{Phys. Rev. E}} \textbf{\bibinfo{volume}{78}},
  \bibinfo{pages}{031923} (\bibinfo{year}{2008}).
\newblock \urlprefix\url{https://link.aps.org/doi/10.1103/PhysRevE.78.031923}.

\bibitem{Hoffman2011}
\bibinfo{author}{Hoffman, B.~D.}, \bibinfo{author}{Grashoff, C.} \&
  \bibinfo{author}{Schwartz, M.~A.}
\newblock \bibinfo{title}{Dynamic molecular processes mediate cellular
  mechanotransduction}.
\newblock \emph{\bibinfo{journal}{Nature}} \textbf{\bibinfo{volume}{475}},
  \bibinfo{pages}{316--323} (\bibinfo{year}{2011}).
\newblock \urlprefix\url{https://doi.org/10.1038/nature10316}.

\bibitem{Iwadate2013}
\bibinfo{author}{Iwadate, Y.} \emph{et~al.}
\newblock \bibinfo{title}{Myosin-ii-mediated directional migration of
  dictyostelium cells in response to cyclic stretching of substratum}.
\newblock \emph{\bibinfo{journal}{Biophysical Journal}}
  \textbf{\bibinfo{volume}{104}}, \bibinfo{pages}{748 -- 758}
  (\bibinfo{year}{2013}).
\newblock
  \urlprefix\url{http://www.sciencedirect.com/science/article/pii/S0006349513000714}.

\bibitem{Cui2015}
\bibinfo{author}{Cui, Y.} \emph{et~al.}
\newblock \bibinfo{title}{Cyclic stretching of soft substrates induces
  spreading and growth}.
\newblock \emph{\bibinfo{journal}{Nature Communications}}
  \textbf{\bibinfo{volume}{6}}, \bibinfo{pages}{6333} (\bibinfo{year}{2015}).
\newblock \urlprefix\url{https://doi.org/10.1038/ncomms7333}.

\bibitem{Gudipaty2017}
\bibinfo{author}{Gudipaty, S.~A.} \emph{et~al.}
\newblock \bibinfo{title}{Mechanical stretch triggers rapid epithelial cell
  division through piezo1}.
\newblock \emph{\bibinfo{journal}{Nature}} \textbf{\bibinfo{volume}{543}},
  \bibinfo{pages}{118--121} (\bibinfo{year}{2017}).
\newblock \urlprefix\url{https://doi.org/10.1038/nature21407}.

\bibitem{Asano2018}
\bibinfo{author}{Asano, S.} \emph{et~al.}
\newblock \bibinfo{title}{Cyclic stretch enhances reorientation and
  differentiation of 3-d culture model of human airway smooth muscle}.
\newblock \emph{\bibinfo{journal}{Biochemistry and Biophysics Reports}}
  \textbf{\bibinfo{volume}{16}}, \bibinfo{pages}{32 -- 38}
  (\bibinfo{year}{2018}).
\newblock
  \urlprefix\url{http://www.sciencedirect.com/science/article/pii/S2405580818301602}.

\bibitem{roca2017quantifying}
\bibinfo{author}{Roca-Cusachs, P.}, \bibinfo{author}{Conte, V.} \&
  \bibinfo{author}{Trepat, X.}
\newblock \bibinfo{title}{Quantifying forces in cell biology}.
\newblock \emph{\bibinfo{journal}{Nature cell biology}}
  \textbf{\bibinfo{volume}{19}}, \bibinfo{pages}{742--751}
  (\bibinfo{year}{2017}).

\bibitem{Merkel2007}
\bibinfo{author}{Merkel, R.}, \bibinfo{author}{Kirchgeßner, N.},
  \bibinfo{author}{Cesa, C.~M.} \& \bibinfo{author}{Hoffmann, B.}
\newblock \bibinfo{title}{Cell force microscopy on elastic layers of finite
  thickness}.
\newblock \emph{\bibinfo{journal}{Biophysical Journal}}
  \textbf{\bibinfo{volume}{93}}, \bibinfo{pages}{3314 -- 3323}
  (\bibinfo{year}{2007}).
\newblock
  \urlprefix\url{http://www.sciencedirect.com/science/article/pii/S0006349507715850}.

\bibitem{maloney2008influence}
\bibinfo{author}{Maloney, J.~M.}, \bibinfo{author}{Walton, E.~B.},
  \bibinfo{author}{Bruce, C.~M.} \& \bibinfo{author}{Van~Vliet, K.~J.}
\newblock \bibinfo{title}{Influence of finite thickness and stiffness on
  cellular adhesion-induced deformation of compliant substrata}.
\newblock \emph{\bibinfo{journal}{Physical Review E}}
  \textbf{\bibinfo{volume}{78}}, \bibinfo{pages}{041923}
  (\bibinfo{year}{2008}).

\bibitem{Sen2009}
\bibinfo{author}{Sen, S.}, \bibinfo{author}{Engler, A.~J.} \&
  \bibinfo{author}{Discher, D.~E.}
\newblock \bibinfo{title}{Matrix strains induced by cells: Computing how far
  cells can feel}.
\newblock \emph{\bibinfo{journal}{Cellular and Molecular Bioengineering}}
  \textbf{\bibinfo{volume}{2}}, \bibinfo{pages}{39--48} (\bibinfo{year}{2009}).
\newblock \urlprefix\url{https://doi.org/10.1007/s12195-009-0052-z}.

\bibitem{Buxboim297}
\bibinfo{author}{Buxboim, A.}, \bibinfo{author}{Ivanovska, I.~L.} \&
  \bibinfo{author}{Discher, D.~E.}
\newblock \bibinfo{title}{Matrix elasticity, cytoskeletal forces and physics of
  the nucleus: how deeply do cells {\textquoteleft}feel{\textquoteright}
  outside and in?} \textbf{\bibinfo{volume}{123}}, \bibinfo{pages}{297--308}
  (\bibinfo{year}{2010}).
\newblock \urlprefix\url{https://jcs.biologists.org/content/123/3/297}.

\bibitem{Bertet2004}
\bibinfo{author}{Bertet, C.}, \bibinfo{author}{Sulak, L.} \&
  \bibinfo{author}{Lecuit, T.}
\newblock \bibinfo{title}{Myosin-dependent junction remodelling controls planar
  cell intercalation and axis elongation}.
\newblock \emph{\bibinfo{journal}{Nature}} \textbf{\bibinfo{volume}{429}},
  \bibinfo{pages}{667--671} (\bibinfo{year}{2004}).
\newblock \urlprefix\url{https://doi.org/10.1038/nature02590}.

\bibitem{Martin2009}
\bibinfo{author}{Martin, A.~C.}, \bibinfo{author}{Kaschube, M.} \&
  \bibinfo{author}{Wieschaus, E.~F.}
\newblock \bibinfo{title}{Pulsed contractions of an actin--myosin network drive
  apical constriction}.
\newblock \emph{\bibinfo{journal}{Nature}} \textbf{\bibinfo{volume}{457}},
  \bibinfo{pages}{495--499} (\bibinfo{year}{2009}).
\newblock \urlprefix\url{https://doi.org/10.1038/nature07522}.

\bibitem{Heisenberg2013}
\bibinfo{author}{Heisenberg, C.-P.} \& \bibinfo{author}{Bella{\"i}che, Y.}
\newblock \bibinfo{title}{Forces in tissue morphogenesis and patterning}.
\newblock \emph{\bibinfo{journal}{Cell}} \textbf{\bibinfo{volume}{153}},
  \bibinfo{pages}{948--962} (\bibinfo{year}{2013}).
\newblock \urlprefix\url{https://doi.org/10.1016/j.cell.2013.05.008}.

\bibitem{Toyama1683}
\bibinfo{author}{Toyama, Y.}, \bibinfo{author}{Peralta, X.~G.},
  \bibinfo{author}{Wells, A.~R.}, \bibinfo{author}{Kiehart, D.~P.} \&
  \bibinfo{author}{Edwards, G.~S.}
\newblock \bibinfo{title}{Apoptotic force and tissue dynamics during drosophila
  embryogenesis} \textbf{\bibinfo{volume}{321}}, \bibinfo{pages}{1683--1686}
  (\bibinfo{year}{2008}).
\newblock \urlprefix\url{https://science.sciencemag.org/content/321/5896/1683}.

\bibitem{Teng2011}
\bibinfo{author}{Teng, X.} \& \bibinfo{author}{Toyama, Y.}
\newblock \bibinfo{title}{Apoptotic force: Active mechanical function of cell
  death during morphogenesis}.
\newblock \emph{\bibinfo{journal}{Development, Growth \& Differentiation}}
  \textbf{\bibinfo{volume}{53}}, \bibinfo{pages}{269--276}.

\bibitem{Trepat2009}
\bibinfo{author}{Trepat, X.} \emph{et~al.}
\newblock \bibinfo{title}{Physical forces during collective cell migration}.
\newblock \emph{\bibinfo{journal}{Nature Physics}}
  \textbf{\bibinfo{volume}{5}}, \bibinfo{pages}{426--430}
  (\bibinfo{year}{2009}).
\newblock \urlprefix\url{https://doi.org/10.1038/nphys1269}.

\bibitem{Tambe2011}
\bibinfo{author}{Tambe, D.~T.} \emph{et~al.}
\newblock \bibinfo{title}{Collective cell guidance by cooperative intercellular
  forces}.
\newblock \emph{\bibinfo{journal}{Nature Materials}}
  \textbf{\bibinfo{volume}{10}}, \bibinfo{pages}{469--475}
  (\bibinfo{year}{2011}).
\newblock \urlprefix\url{https://doi.org/10.1038/nmat3025}.

\bibitem{Aoki2017}
\bibinfo{author}{Aoki, K.} \emph{et~al.}
\newblock \bibinfo{title}{Propagating wave of erk activation orients collective
  cell migration}.
\newblock \emph{\bibinfo{journal}{Developmental Cell}}
  \textbf{\bibinfo{volume}{43}}, \bibinfo{pages}{305 -- 317.e5}
  (\bibinfo{year}{2017}).
\newblock
  \urlprefix\url{http://www.sciencedirect.com/science/article/pii/S1534580717308298}.

\bibitem{Hino2020}
\bibinfo{author}{Hino, N.} \emph{et~al.}
\newblock \bibinfo{title}{Erk-mediated mechanochemical waves direct collective
  cell polarization}.
\newblock \emph{\bibinfo{journal}{Developmental Cell}}
  \textbf{\bibinfo{volume}{53}}, \bibinfo{pages}{646 -- 660.e8}
  (\bibinfo{year}{2020}).
\newblock
  \urlprefix\url{http://www.sciencedirect.com/science/article/pii/S1534580720304007}.

\bibitem{boocock2021theory}
\bibinfo{author}{Boocock, D.}, \bibinfo{author}{Hino, N.},
  \bibinfo{author}{Ruzickova, N.}, \bibinfo{author}{Hirashima, T.} \&
  \bibinfo{author}{Hannezo, E.}
\newblock \bibinfo{title}{Theory of mechanochemical patterning and optimal
  migration in cell monolayers}.
\newblock \emph{\bibinfo{journal}{Nature Physics}}
  \textbf{\bibinfo{volume}{17}}, \bibinfo{pages}{267--274}
  (\bibinfo{year}{2021}).

\bibitem{Fukuyama2020arXiv}
\bibinfo{author}{Fukuyama, T.}, \bibinfo{author}{Ebata, H.},
  \bibinfo{author}{Kondo, Y.}, \bibinfo{author}{Aoki, K.} \&
  \bibinfo{author}{Maeda, Y.~T.}
\newblock \bibinfo{title}{Why epithelial cells collectively move against a
  traveling signal wave} (\bibinfo{year}{2020}).
\newblock \urlprefix\url{https://arxiv.org/abs/2008.12955"}.
\newblock \bibinfo{note}{Preprint}.

\bibitem{Uroz2018}
\bibinfo{author}{Uroz, M.} \emph{et~al.}
\newblock \bibinfo{title}{Regulation of cell cycle progression by cell--cell
  and cell--matrix forces}.
\newblock \emph{\bibinfo{journal}{Nature Cell Biology}}
  \textbf{\bibinfo{volume}{20}}, \bibinfo{pages}{646--654}
  (\bibinfo{year}{2018}).
\newblock \urlprefix\url{https://doi.org/10.1038/s41556-018-0107-2}.

\bibitem{zhou2015biotribology}
\bibinfo{author}{Zhou, Z.} \& \bibinfo{author}{Jin, Z.}
\newblock \bibinfo{title}{Biotribology: recent progresses and future
  perspectives}.
\newblock \emph{\bibinfo{journal}{Biosurface and biotribology}}
  \textbf{\bibinfo{volume}{1}}, \bibinfo{pages}{3--24} (\bibinfo{year}{2015}).

\bibitem{barnett1962lubrication}
\bibinfo{author}{Barnett, C.~H.} \& \bibinfo{author}{Cobbold, A.}
\newblock \bibinfo{title}{Lubrication within living joints}.
\newblock \emph{\bibinfo{journal}{The Journal of Bone and Joint Surgery.
  British volume}} \textbf{\bibinfo{volume}{44}}, \bibinfo{pages}{662--674}
  (\bibinfo{year}{1962}).

\bibitem{shacham2010measurements}
\bibinfo{author}{Shacham, S.}, \bibinfo{author}{Castel, D.} \&
  \bibinfo{author}{Gefen, A.}
\newblock \bibinfo{title}{Measurements of the static friction coefficient
  between bone and muscle tissues}.
\newblock \emph{\bibinfo{journal}{J. Biomech. Eng.}}
  \textbf{\bibinfo{volume}{132}}, \bibinfo{pages}{084502}
  (\bibinfo{year}{2010}).

\bibitem{merkher2006rational}
\bibinfo{author}{Merkher, Y.} \emph{et~al.}
\newblock \bibinfo{title}{A rational human joint friction test using a human
  cartilage-on-cartilage arrangement}.
\newblock \emph{\bibinfo{journal}{Tribology Letters}}
  \textbf{\bibinfo{volume}{22}}, \bibinfo{pages}{29--36}
  (\bibinfo{year}{2006}).

\bibitem{katta2008biotribology}
\bibinfo{author}{Katta, J.}, \bibinfo{author}{Jin, Z.},
  \bibinfo{author}{Ingham, E.} \& \bibinfo{author}{Fisher, J.}
\newblock \bibinfo{title}{Biotribology of articular cartilage—a review of the
  recent advances}.
\newblock \emph{\bibinfo{journal}{Medical engineering \& physics}}
  \textbf{\bibinfo{volume}{30}}, \bibinfo{pages}{1349--1363}
  (\bibinfo{year}{2008}).

\bibitem{kawaue}
\bibinfo{author}{Kawaue, T.} \emph{et~al.}
\newblock \bibinfo{title}{Mechanics defines the spatial pattern of compensatory
  proliferation}.
\newblock \bibinfo{note}{Preprint}.

\bibitem{leong2015viscoelastic}
\bibinfo{author}{Leong, M.~C.}, \bibinfo{author}{Nai, M.~H.},
  \bibinfo{author}{Cheong, F.~C.}, \bibinfo{author}{Lim, C.~T.} \emph{et~al.}
\newblock \bibinfo{title}{Viscoelastic effects of silicone gels at the
  micro-and nanoscale}.
\newblock \emph{\bibinfo{journal}{Procedia IUTAM}}
  \textbf{\bibinfo{volume}{12}}, \bibinfo{pages}{20--30}
  (\bibinfo{year}{2015}).

\bibitem{Teng2017}
\bibinfo{title}{Remodeling of adhesion and modulation of mechanical tensile
  forces during apoptosis in drosophila epithelium}.
\newblock \emph{\bibinfo{journal}{Development}} \textbf{\bibinfo{volume}{144}},
  \bibinfo{pages}{95--105} (\bibinfo{year}{2017}).

\bibitem{wang2002cell}
\bibinfo{author}{Wang, N.} \emph{et~al.}
\newblock \bibinfo{title}{Cell prestress. i. stiffness and prestress are
  closely associated in adherent contractile cells}.
\newblock \emph{\bibinfo{journal}{American Journal of Physiology-Cell
  Physiology}} \textbf{\bibinfo{volume}{282}}, \bibinfo{pages}{C606--C616}
  (\bibinfo{year}{2002}).

\bibitem{kruse2006contractility}
\bibinfo{author}{Kruse, K.}, \bibinfo{author}{Joanny, J.},
  \bibinfo{author}{J{\"u}licher, F.} \& \bibinfo{author}{Prost, J.}
\newblock \bibinfo{title}{Contractility and retrograde flow in lamellipodium
  motion}.
\newblock \emph{\bibinfo{journal}{Physical biology}}
  \textbf{\bibinfo{volume}{3}}, \bibinfo{pages}{130} (\bibinfo{year}{2006}).

\bibitem{serra2012mechanical}
\bibinfo{author}{Serra-Picamal, X.} \emph{et~al.}
\newblock \bibinfo{title}{Mechanical waves during tissue expansion}.
\newblock \emph{\bibinfo{journal}{Nature Physics}}
  \textbf{\bibinfo{volume}{8}}, \bibinfo{pages}{628--634}
  (\bibinfo{year}{2012}).

\bibitem{deforet2014emergence}
\bibinfo{author}{Deforet, M.}, \bibinfo{author}{Hakim, V.},
  \bibinfo{author}{Yevick, H.}, \bibinfo{author}{Duclos, G.} \&
  \bibinfo{author}{Silberzan, P.}
\newblock \bibinfo{title}{Emergence of collective modes and tri-dimensional
  structures from epithelial confinement}.
\newblock \emph{\bibinfo{journal}{Nature communications}}
  \textbf{\bibinfo{volume}{5}}, \bibinfo{pages}{1--9} (\bibinfo{year}{2014}).

\bibitem{blanch2017hydrodynamic}
\bibinfo{author}{Blanch-Mercader, C.} \& \bibinfo{author}{Casademunt, J.}
\newblock \bibinfo{title}{Hydrodynamic instabilities, waves and turbulence in
  spreading epithelia}.
\newblock \emph{\bibinfo{journal}{Soft matter}} \textbf{\bibinfo{volume}{13}},
  \bibinfo{pages}{6913--6928} (\bibinfo{year}{2017}).

\bibitem{lu2012extracellular}
\bibinfo{author}{Lu, P.}, \bibinfo{author}{Weaver, V.~M.} \&
  \bibinfo{author}{Werb, Z.}
\newblock \bibinfo{title}{The extracellular matrix: a dynamic niche in cancer
  progression}.
\newblock \emph{\bibinfo{journal}{Journal of cell biology}}
  \textbf{\bibinfo{volume}{196}}, \bibinfo{pages}{395--406}
  (\bibinfo{year}{2012}).

\bibitem{TAWADA1991193}
\bibinfo{author}{Tawada, K.} \& \bibinfo{author}{Sekimoto, K.}
\newblock \bibinfo{title}{Protein friction exerted by motor enzymes through a
  weak-binding interaction}.
\newblock \emph{\bibinfo{journal}{Journal of Theoretical Biology}}
  \textbf{\bibinfo{volume}{150}}, \bibinfo{pages}{193--200}
  (\bibinfo{year}{1991}).
\newblock
  \urlprefix\url{https://www.sciencedirect.com/science/article/pii/S0022519305803315}.

\bibitem{panzetta2019cell}
\bibinfo{author}{Panzetta, V.}, \bibinfo{author}{Fusco, S.} \&
  \bibinfo{author}{Netti, P.~A.}
\newblock \bibinfo{title}{Cell mechanosensing is regulated by substrate strain
  energy rather than stiffness}.
\newblock \emph{\bibinfo{journal}{Proceedings of the National Academy of
  Sciences}} \textbf{\bibinfo{volume}{116}}, \bibinfo{pages}{22004--22013}
  (\bibinfo{year}{2019}).

\bibitem{dupont2011role}
\bibinfo{author}{Dupont, S.} \emph{et~al.}
\newblock \bibinfo{title}{Role of yap/taz in mechanotransduction}.
\newblock \emph{\bibinfo{journal}{Nature}} \textbf{\bibinfo{volume}{474}},
  \bibinfo{pages}{179--183} (\bibinfo{year}{2011}).

\bibitem{nardone2017yap}
\bibinfo{author}{Nardone, G.} \emph{et~al.}
\newblock \bibinfo{title}{Yap regulates cell mechanics by controlling focal
  adhesion assembly}.
\newblock \emph{\bibinfo{journal}{Nature communications}}
  \textbf{\bibinfo{volume}{8}}, \bibinfo{pages}{1--13} (\bibinfo{year}{2017}).

\bibitem{KOCGOZLU20162942}
\bibinfo{author}{Kocgozlu, L.} \emph{et~al.}
\newblock \bibinfo{title}{Epithelial cell packing induces distinct modes of
  cell extrusions}.
\newblock \emph{\bibinfo{journal}{Current Biology}}
  \textbf{\bibinfo{volume}{26}}, \bibinfo{pages}{2942--2950}
  (\bibinfo{year}{2016}).
\newblock
  \urlprefix\url{https://www.sciencedirect.com/science/article/pii/S0960982216310028}.

\bibitem{RevModPhys.85.1143}
\bibinfo{author}{Marchetti, M.~C.} \emph{et~al.}
\newblock \bibinfo{title}{Hydrodynamics of soft active matter}.
\newblock \emph{\bibinfo{journal}{Rev. Mod. Phys.}}
  \textbf{\bibinfo{volume}{85}}, \bibinfo{pages}{1143--1189}
  (\bibinfo{year}{2013}).
\newblock \urlprefix\url{https://link.aps.org/doi/10.1103/RevModPhys.85.1143}.

\bibitem{Maruthamuthu4708}
\bibinfo{author}{Maruthamuthu, V.}, \bibinfo{author}{Sabass, B.},
  \bibinfo{author}{Schwarz, U.~S.} \& \bibinfo{author}{Gardel, M.~L.}
\newblock \bibinfo{title}{Cell-ecm traction force modulates endogenous tension
  at cell{\textendash}cell contacts} \textbf{\bibinfo{volume}{108}},
  \bibinfo{pages}{4708--4713} (\bibinfo{year}{2011}).
\newblock \urlprefix\url{https://www.pnas.org/content/108/12/4708}.

\bibitem{Liu9944}
\bibinfo{author}{Liu, Z.} \emph{et~al.}
\newblock \bibinfo{title}{Mechanical tugging force regulates the size of
  cell{\textendash}cell junctions} \textbf{\bibinfo{volume}{107}},
  \bibinfo{pages}{9944--9949} (\bibinfo{year}{2010}).
\newblock \urlprefix\url{https://www.pnas.org/content/107/22/9944}.

\bibitem{Khalilgharibi2019}
\bibinfo{author}{Khalilgharibi, N.} \emph{et~al.}
\newblock \bibinfo{title}{Stress relaxation in epithelial monolayers is
  controlled by the actomyosin cortex}.
\newblock \emph{\bibinfo{journal}{Nature Physics}}
  \textbf{\bibinfo{volume}{15}}, \bibinfo{pages}{839--847}
  (\bibinfo{year}{2019}).
\newblock \urlprefix\url{https://www.nature.com/articles/s41567-019-0516-6}.

\bibitem{fletcher2010cell}
\bibinfo{author}{Fletcher, D.~A.} \& \bibinfo{author}{Mullins, R.~D.}
\newblock \bibinfo{title}{Cell mechanics and the cytoskeleton}.
\newblock \emph{\bibinfo{journal}{Nature}} \textbf{\bibinfo{volume}{463}},
  \bibinfo{pages}{485--492} (\bibinfo{year}{2010}).

\bibitem{rossen2014long}
\bibinfo{author}{Rossen, N.~S.}, \bibinfo{author}{Tarp, J.~M.},
  \bibinfo{author}{Mathiesen, J.}, \bibinfo{author}{Jensen, M.~H.} \&
  \bibinfo{author}{Oddershede, L.~B.}
\newblock \bibinfo{title}{Long-range ordered vorticity patterns in living
  tissue induced by cell division}.
\newblock \emph{\bibinfo{journal}{Nature communications}}
  \textbf{\bibinfo{volume}{5}}, \bibinfo{pages}{1--7} (\bibinfo{year}{2014}).

\bibitem{charras2018tensile}
\bibinfo{author}{Charras, G.} \& \bibinfo{author}{Yap, A.~S.}
\newblock \bibinfo{title}{Tensile forces and mechanotransduction at cell--cell
  junctions}.
\newblock \emph{\bibinfo{journal}{Current Biology}}
  \textbf{\bibinfo{volume}{28}}, \bibinfo{pages}{R445--R457}
  (\bibinfo{year}{2018}).

\end{thebibliography}

\bibliographystyle{naturemag}

\section*{Acknowledgments}
We thank Andrew Wong from Mechanobiology Institute (MBI) science communication core  for editing the manuscript and MBI computational core for supporting us about computer-related research activities. We also appreciate Jennifer Young, Rakesh Das, Alokendra Ghosh (MBI) and Ayumi Ozawa (University of Tokyo) for valuable discussions.
This research was supported by Singapore Ministry of Education Tier 2 grant, MOE2015-T2-1-116, MOE2020-T2-2-033 (to YT),
Japan Society for the Promotion of Science (JSPS) Overseas Research Fellowships (to TK),
and Seed fund of Mechanobiology Institute (to YT, JP, TH).

%\subsection*{Funding}
%Singapore Ministry of Education Tier 2 grant, MOE2015-T2-1-116, MOE2020-T2-2-033 (YT)\\
%Japan Society for the Promotion of Science (JSPS) Overseas Research Fellowships (TK)\\
%Mechanobiology Institute, National University of Singapore, Seed fund (YT, JP, TH).

\subsection*{Author contributions}
YL, YT, JP and TH conceptualized the work, 
TK and IY performed experiments, 
YL performed data analysis, 
YL, JP and TH constructed theory,
YL visualized the results,
TH, JP, and YT supervised, 
YL and TH wrote the original draft, and 
YL, TH, JP and YT reviewed and edited. 
%Conceptualization: YL, YT, JP and TH \\ 
%Experiment: TK and IY \\
%Data analysis: YL\\
%Theory: YL, JP and TH\\  
%Visualization: YL \\ 
%Supervision: TH, JP, and YT\\
%Writing--original draft: YL and TH\\
%Writing--review\&editing: JP and YT

\subsection*{Competing interests}
All other authors declare they have no competing interests.
\subsection*{Data and availability}
All data are available in the main text or the supplementary materials.
\clearpage
\begin{figure}
\centering
\includegraphics[width=0.85\linewidth]{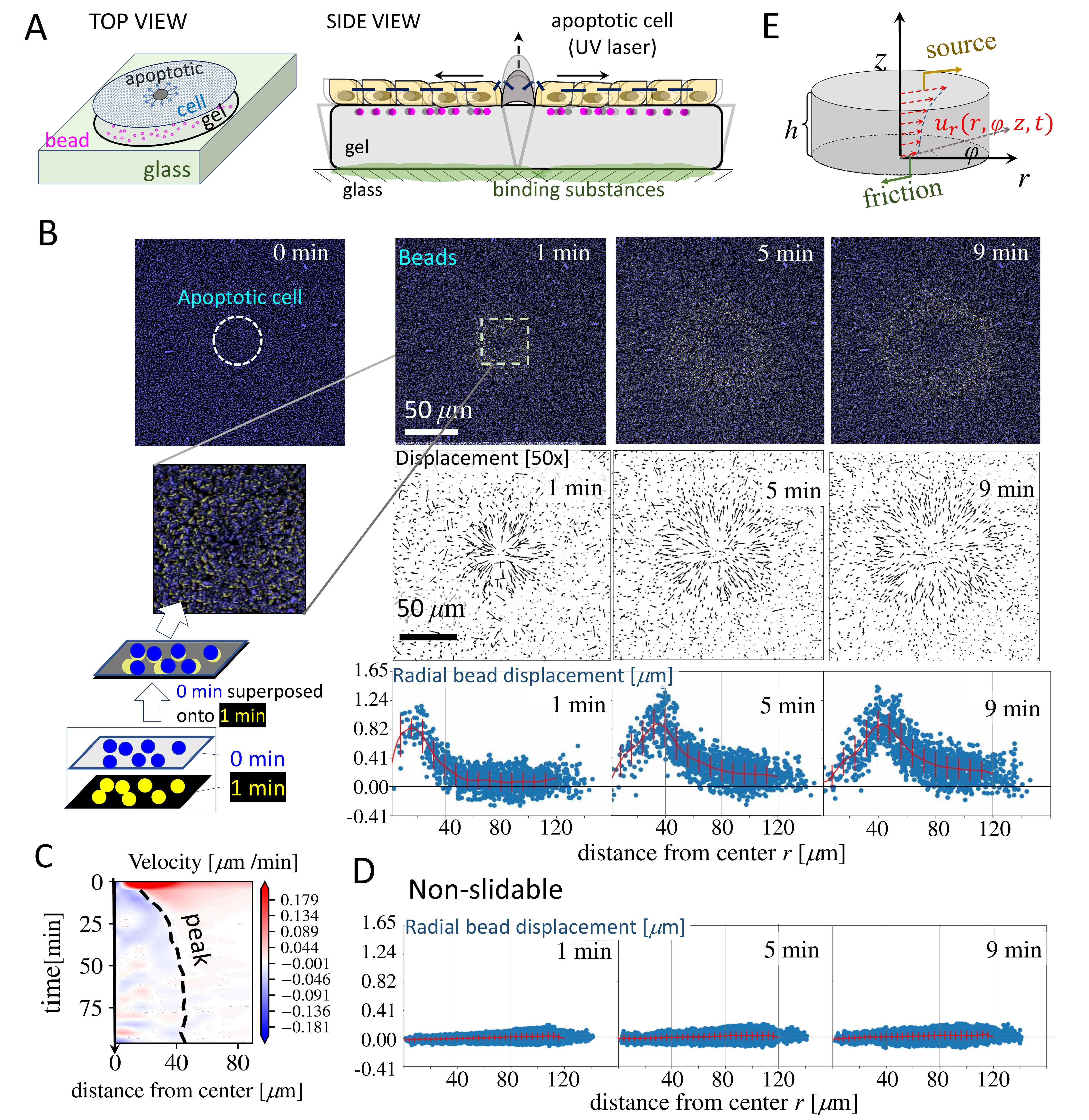}
\caption{Time-dependent traction force microscopy (TFM) setup for studying the interfacial effect of substrate layers. (A) Cell apoptosis provides the force origin. The amount of binding substances between the gel and glass is controlled, which is different from a conventional TFM. (B) Concentric waves emerge with a weakly adhesive gel-glass interface. Top: images of florescent beads under TFM; Color of beads were tuned as blue for $t=0$ min, and as yellow for $t>0$ min. The snapshots $t>0$ min were superposed onto the snapshot of $t=0$ min for visualizing the realistic bead displacement. Middle: vector fields of bead displacement by tracking the beads in the images (arrows length 50 times magnified). Bottom: displacement of beads by projecting the vector field of displacement onto radial direction (blue dots). The red curves are the smoothed moving average of the scattered dots. (C) A typical Kymograph for smoothed displacement velocity. The dashed curve represents where the peak of radial displacement locates. (D) Radial displacement of beads with the gel strongly bonded to the glass (non-slidable). (E) 3D elastic material model for the deformation problem in the gel in our experiments.}\label{fig:scheme}
\end{figure}

\begin{figure}
\centering
\includegraphics[width=\linewidth]{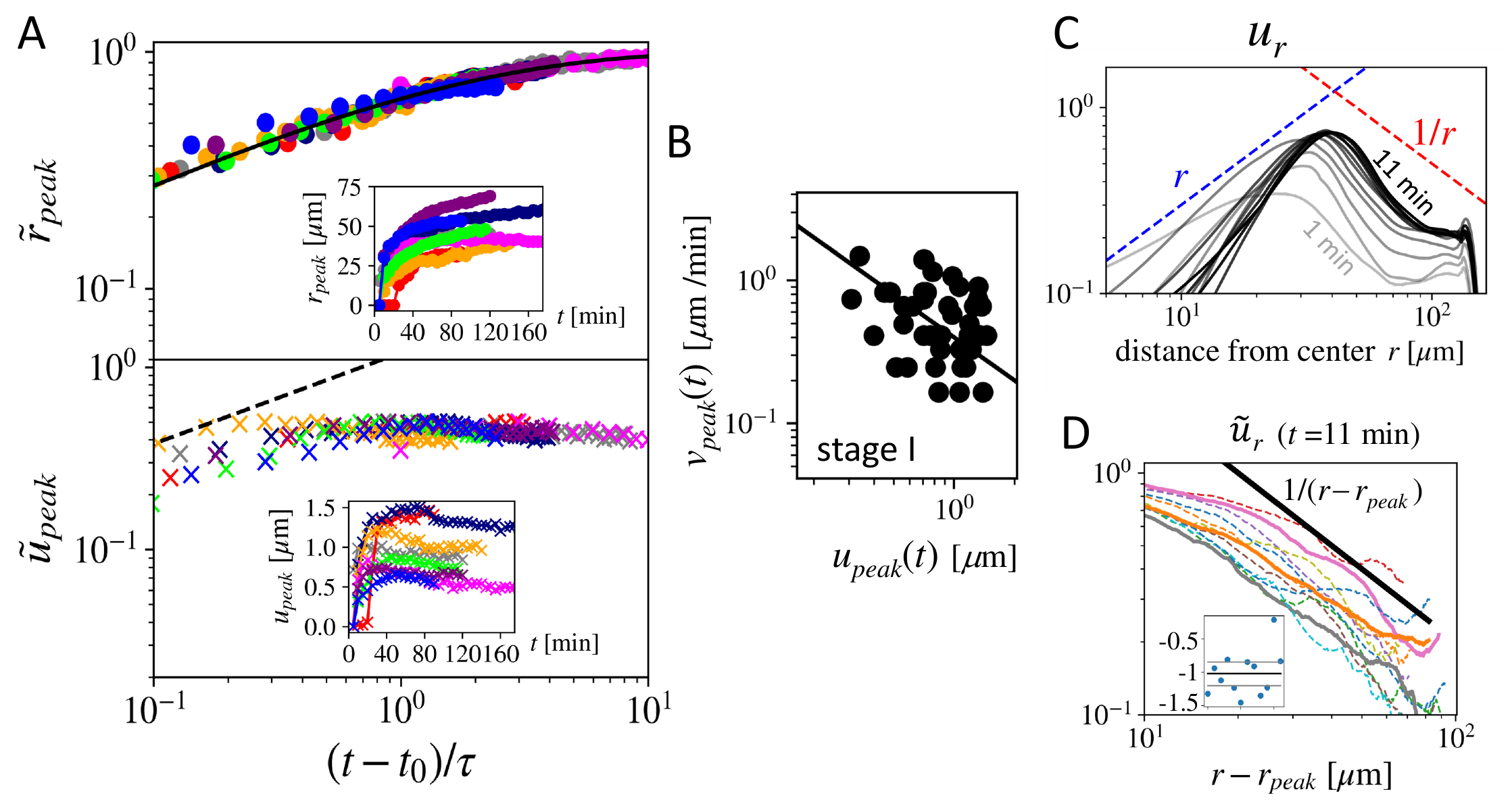}
\caption{Substrate dynamics in experiment. (A) The dynamics of the rescaled peak position $\tilde{r}_{peak}$ and rescaled peak magnitude $\tilde{u}_{peak}$. The black solid curve in the top plot shows the fitting form (Eq.\ref{eq:fitting} in Materials and Methods) where a diffusive trend crosses over to a saturating plateau; the black dashed curve in the bottom plot shows a diffusive trend. Two stages in time are separated by the onset of the deviation of $\tilde{u}_{peak}$ from a diffusive behavior. Inset: Original sample-wise peak dynamics. (B) The negative correlation between peak propagation speed $v_{peak}$ and peak magnitude $u_{peak}$ for all the time points within stage I ($t<\tau$). The straight line corresponds to $v_{peak}=0.4/u_{peak}$. (C) Evolution of smoothed moving average of beads radial displacement $u_r$ over time from $1$ min (light gray) to 11 min (black) in a typical sample. (D) Rescaled radial displacement $\tilde{u}$ of multiple samples 11 minutes after apoptosis for the relative distance from peak $r-r_{peak}$. The bold black line corresponds to $\propto 1/(r-r_{peak})$. Inset shows the fitted tail exponents (blue dots) and the bold black line is the average over samples with the two gray lines sketching a 90\% confidence interval $-1.00\pm 0.18$. \label{fig:peak}}
\end{figure}

\begin{figure}
\centering
\includegraphics[width=1.0\linewidth]{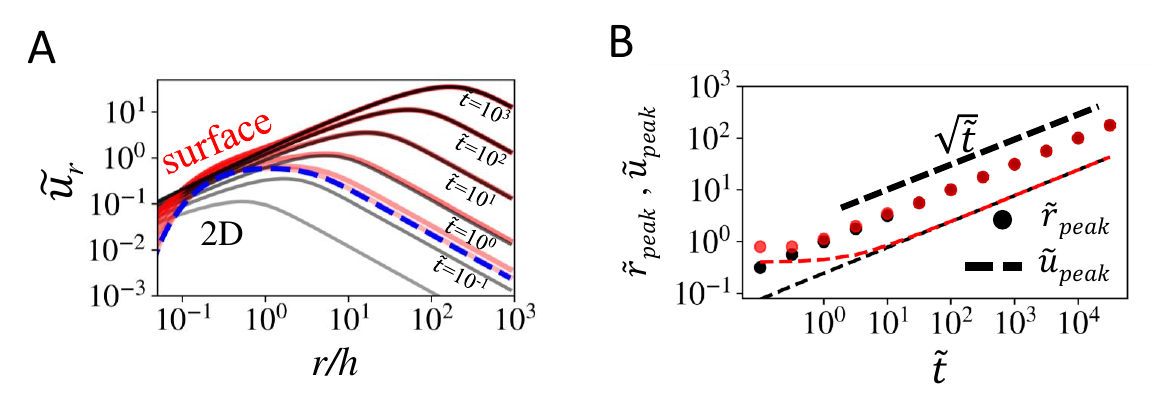} 
\caption{Emergence of peak propagation in theory: solutions with the simple surface stress form given by Eq.~\ref{eq:spreading} with $q=1$. (A) Radial displacement $\tilde{u_r}$, which is $u_r$ normalized by a coefficient $s\varepsilon/h\hat{G}$ at near surface z=0.9h (red) and under 2D approximation (black) for varying dimensionless time $\tilde{t}=t/t_c=t\hat{G}/\xi h$. (B): Dynamics of $\tilde{r}_{peak}$, which is $r_{peak}$ normalized by $h$ (dotted lines) and $\tilde{u}_{peak}$, which $u_{peak}$ normalized by $s\varepsilon/h\hat{G}$(dashed lines) for the near-surface (red) and 2D approximation model with dimensionless time $\tilde{t}$. The corresponding results for the cases with more comprehensive forms of surface stress fields (see Sections S2.2 and S2.3) are provided in Fig.~S3.  \label{fig:theory}}
\end{figure}

\begin{figure}
\centering
\includegraphics[width=0.7\linewidth]{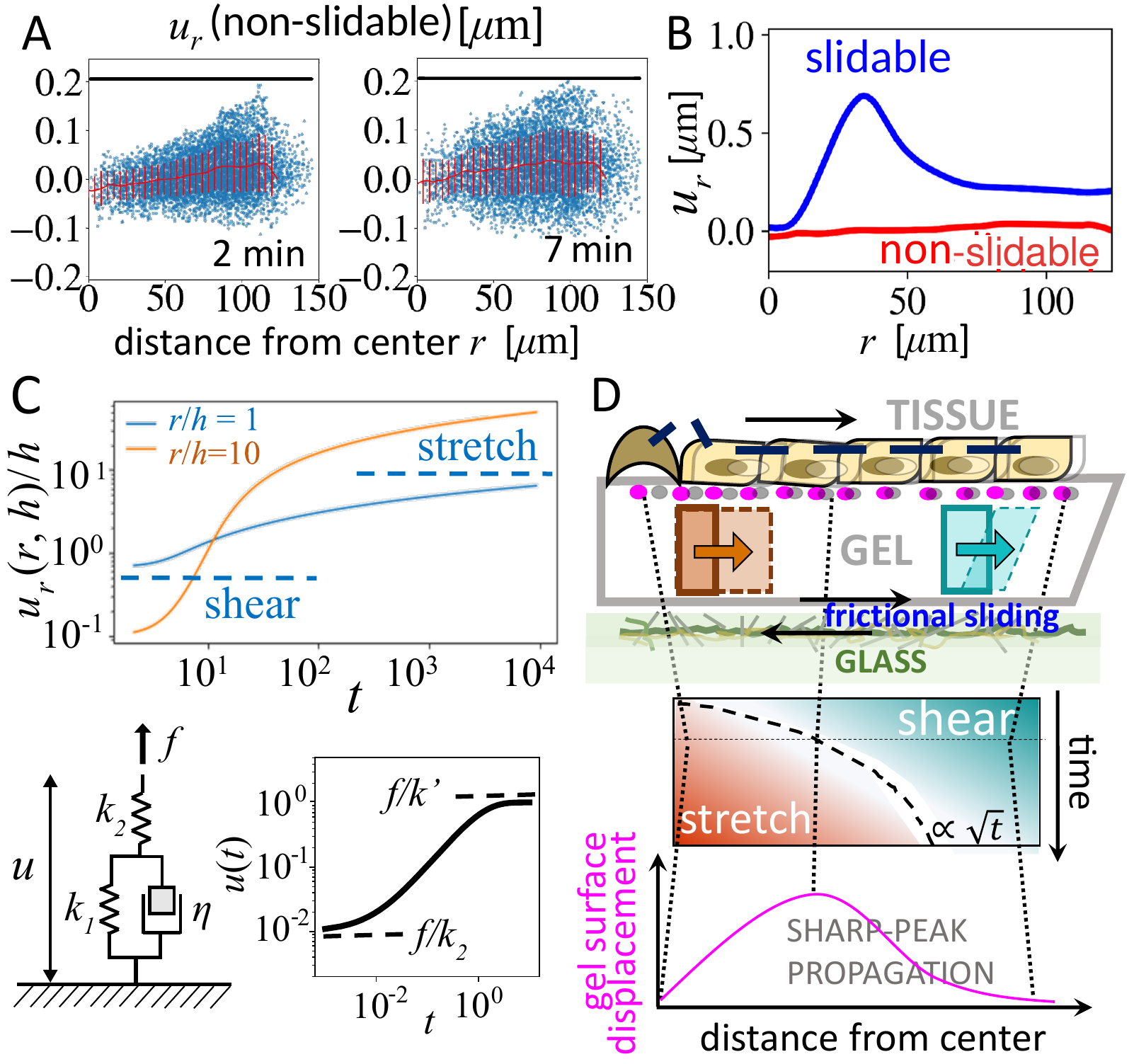} 
\caption{Illustration of the physics underlying the deformation propagation in substrate. (A) Absence of propagation with a strongly adhesive (non-slidable) gel base. Black line indicates the resolution threshold at $0.206\mu$m. (B) A comparison between the two typical samples under slidable (blue) and non-slidable (red) boundary conditions. (C) Top:  ``Creeping" dynamics of effective strain $u_r/h$ (normalized by $s\varepsilon/h\hat{G}$) at the gel surface for near field (blue curve) and far field (orange curve).
Bottom: Creeping curve of a type of viscoelasticity model (Zener model) described in the schematic illustration. $f$ represents the external stress, $k_1$ and $k_2$ mean two different parts of elastic modulus, $\eta$ is the viscosity. Time evolution of the deformation $u$ is given by $u(t)=(f/k_1)[1-\exp(-t/\tau)] + f/k_2$ with $\tau=\eta/k_1$. Effective modulus $k'=1/(1/k_1+1/k_2)$. For the graph, the parameters are set $f=1$, $k_1=1$, $\eta=1$ (units) and $k_2=100$.
(D) Pedagogical picture of propagation in our 3-layer system. Frictional sliding at the gel-glass interface causes a diffusive expansion of the stretch regime from the center ($r=0$) in the gel over time. Under a spatially decaying form of surface stress $S(r,t)$, the gel deforms mostly at the boundary between the stretch- and the shear-dominated regions. The kymograph of deformation type at the middle is reminiscent of the kymograph of beads velocity in Fig. \ref{fig:scheme}C; we have proposed the reason that the stretching deformation will diminish (gel moving inward) as the stress exerted by the tissue decays with time.} \label{fig:disc}
\end{figure}

\begin{table}
\centering
    \caption{\label{info}Summary of experiment information.}
    \begin{tabular}{ccccccc}
   No. & Base & Cell & Interval & Duration &Sample \\
    \hline
   1& slidable & MDCK & 1 min & 11 min & 12\\
   2& slidable & MKN28& 5 min & 2 hours & 8 \\
   3& non-slidable & MDCK & 1 min & 11 min & 3\\
   \hline
    \end{tabular}
\end{table}

%Here you should list the contents of your Supplementary Materials -- below is an example. 
%You should include a list of Supplementary figures, Tables, and any references that appear only in the SM. 
%Note that the reference numbering continues from the main text to the SM.
% In the example below, Refs. 4-10 were cited only in the SM.  
\clearpage
\section*{Supplementary materials}
\setcounter{equation}{0}
\renewcommand{\theequation}{S\arabic{equation}}

\subsection*{S1  3D gel model}
\label{SM:3DGelModel}
Let us consider a 3D vector field in a cylindrical system for the gel displacement 
\begin{equation}
\hat{u}=u_r\hat{e_r}+u_\varphi \hat{e_\varphi} + u_z\hat{e_z},
\label{dis}
\end{equation}
where $u_r,u_\varphi,u_z$ are the scalar field in a 3D cylindrical system. 
The forces transmitted from the tissue layer is represented by a stress field $s(r,\varphi,z=h)$, and this provides a shearing stress at the surface of the gel and leads to the deformation in the gel characterized by the vector displacement field $\hat{u}$. Considering the axisymmetry of the exerted stress field, we have:
\begin{equation}
u_\varphi=0, \ \partial u_r/\partial \varphi=0, \ 
\partial u_z/\partial \varphi=0.
\label{constrain}
\end{equation} 

Based on linear elasticity theory, the strain 
\begin{equation}
\epsilon=\frac{1}{2}[\nabla \hat{u} + (\nabla \hat{u})^T]
\label{let}
\end{equation}
is a second-order symmetric tensor bearing six non-zero components. Considering the symmetry (Eq.\ref{constrain}), the strain tensor in our problem is further simplified to 
\begin{equation}
    \epsilon=
    \begin{pmatrix}
    \epsilon_{rr} & 0 & \epsilon_{rz}\\
    0 & \epsilon_{\varphi \varphi} & 0 \\
    \epsilon_{rz} & 0 & \epsilon_{zz}
    \end{pmatrix}
    \to
    \begin{pmatrix}
        \epsilon_{rr}\\
        \epsilon_{zz}\\
        \epsilon_{\varphi \varphi}\\
        \epsilon_{rz}
    \end{pmatrix},
    \end{equation}
where
\begin{equation}
    \begin{aligned}
   &\epsilon_{rr}=\frac{\partial{u_r}}{\partial{r}},
    \epsilon_{\varphi \varphi}= \frac{u_r}{r}, \\
    &\epsilon_{zz}= \frac{\partial u_z}{\partial z},\ \epsilon_{rz}=\frac{1}{2}\left(\frac{\partial u_r}{\partial z}+\frac{\partial u_z}{\partial r}\right)
    \end{aligned}
    \label{strain}
\end{equation}
are the normal and shear strains with respect to $u_r$ and $u_z$ calculated from Eq.~\ref{let}.

The stress tensor $\sigma$ is related to the strain tensor $\epsilon$ by a stiffness tensor $\hat{E}$:
\begin{equation}
\sigma=\hat{E}\epsilon,
\end{equation}
and in a homogeneous and isotropic material with axisymmetry and Poisson ratio near 0.5 (incompressibility), the stress tensor components simply satisfy
\begin{equation}
    \begin{aligned}
        \sigma_{rr} &= \frac{E}{1+\nu}\epsilon_{rr} - P \ , \ \sigma_{\varphi \varphi} = \frac{E}{1+\nu}\epsilon_{\varphi \varphi}  - P \\
        \sigma_{zz} &= \frac{E}{1+\nu}\epsilon_{zz}  - P \ , \ 
        \sigma_{rz} = \frac{E}{1+\nu}\epsilon_{rz}
    \end{aligned}
    \label{stress-full}
    \end{equation}
with the pressure $P$ being a Lagrange multiplier allowing the deformations to satisfy the incompressibility condition.

The force is balanced in the material as 
\begin{equation}
\nabla \cdot \sigma + \hat{F} = \rho \frac{\partial^2 \hat{u}}{\partial t^2}, 
\end{equation}
where $\sigma$ is the stress tensor and $\hat{F}$ is the body force field and the right-hand term is the inertia. In a micro-scale material, the inertia term is negligible and there is no body force in our problem. Hence, we apply the force balance equation 
\begin{equation}
\nabla \cdot \sigma=0, 
\end{equation}
and again due to the axisymmetry, the force balance along the radial direction is:
\begin{equation}
    \frac{1}{r}\frac{\partial}{\partial r}(r\sigma_{rr})+\frac{\partial\sigma_{rz}}{\partial z}-\frac{\sigma_{\varphi \varphi}}{r}=0,
\label{force}
\end{equation}
and the force balance along the $z$-direction is:
\begin{equation}
    \frac{\partial }{r\partial r}(r\sigma_{rz})+\frac{\partial \sigma_{zz}}{\partial z}=0.
    \label{force-z}
\end{equation}
In what follows, we assume that $\nu$ is asymptotically close to $1/2$, i.e., an incompressible gel.
In this case, we can apply the incompressible condition
\begin{equation}
    \nabla\cdot \hat{u}=\frac{1}{r}\frac{\partial }{\partial r}(r u_r)+\frac{\partial u_z}{\partial z}=\epsilon_{rr}+\epsilon_{\varphi\varphi}+\epsilon_{zz} = 0. \label{eq:incompressibility} 
\end{equation}
to determine the pressure field $P(r,z)$.

In terms of radial displacement field $u_r$, vertical displacement field $u_z$ and pressure field $P$, 
Eqs.\ref{force}-\ref{eq:incompressibility} lead to the following equations :
\begin{equation}
    \frac{\partial^2 u_r}{\partial r^2} + \frac{1}{r}\frac{\partial u_r}{\partial r} - \frac{u_r}{r^2} + \frac{\partial^2 u_r}{\partial z^2} 
     - \frac{2}{\hat{G}} \frac{\partial P}{\partial r} = 0
\label{ur_3d}
\end{equation}
(Eq.\ref{eq:ur_3d} of Section ``Theoretical model and results"),
\begin{equation}
    \frac{\partial^2 u_z}{\partial r^2} + \frac{1}{r}\frac{\partial u_z}{\partial r}  + \frac{\partial^2 u_z}{\partial z^2}
    - \frac{2}{\hat{G}} \frac{\partial P}{\partial z}= 0.
\label{uz_3d}
\end{equation}
and
\begin{equation}
    \frac{\partial^2 P}{\partial r^2} + \frac{1}{r}\frac{\partial P}{\partial r} + \frac{\partial^2 P}{\partial z^2} = 0,
\label{P_3d}
\end{equation}
respectively, where $\hat{G}\equiv E/(1+\nu)=2G$
is 2 times the shear modulus $G$ of the gel.  

\subsubsection*{S1.1  General solution}
\label{Sub}
Here, we provide the solution to Eqs.\ref{ur_3d} - \ref{P_3d} with the following boundary conditions: Two shearing forces on the top and bottome surface (see Eqs.\ref{eq:boundary} in the main text), zero normal forces to the surface $\sigma_{zz}(z=h)=0$, a flat bottom $u_z(z=0)=0$, together with the finiteness of $u_r(r \rightarrow 0)$, $P(r \rightarrow  0)$, $u_r(r \rightarrow \infty)$ and $P(r \rightarrow \infty)$.

Note that $u_r$ and $u_z$ are not independent with each other due to the incompressibility condition
\begin{equation}
    \nabla\cdot \hat{u}=\frac{1}{r}\frac{\partial }{\partial r}(r u_r)+\frac{\partial u_z}{\partial z} = 0 \label{eq:incompressibility2} \ .
\end{equation}

To validate the zero external normal force boundary condition at the gel's top surface, $\sigma_{zz}|_{z=h}=0$, here we roughly compare the gel's elastic normal stress $\sigma_{zz}|_{z=h}^{\rm el}\sim  E\epsilon_{zz} = E\delta h / h$ and tissue tension-originated one $\sigma_{zz}|_{z=h}^{\rm tt}=\gamma \kappa=\gamma \nabla^2 h$ with the gel thickness $h\sim 50 \times 10^{-6} {\rm m}$, the gel's elastic modulus $E \sim 10^4 \mathrm{Pa}$ (used in our experiment) and the tissue linear tension $\gamma = 10^{-3}$N/m considering a 1 $\mu$m-thin belt of junctions  \cite{RevModPhys.85.1143,Maruthamuthu4708,Liu9944,Khalilgharibi2019}. $\delta h$ is the small change in the gel height as compared to the original gel height $h$, and $\kappa= \nabla^2 h$ is the the curvature of the gel surface or the tissue layer, which could be approximated on the scale of cell size $\ell$ as $\delta_c h/\ell^2$, where $\delta_c h/l$ is the gradient of the surface height. One can see it obviously that $\delta_c h$ is smaller than $\delta h$ in our problem when the gel is stretched in the radial direction due to the cell stress.
Eventually, by approximating a typical cell length $\ell \sim 10^{-5} {\mathrm{m}}$, we can compare two normal stresses as $(\sigma_{zz}|_{z=h}^{\mathrm{e}l} )/ (\sigma_{zz}|_{z=h}^{\mathrm{tt}}) = (E \ell^2) / (\gamma h)\times (\delta h/\delta_c h) \sim 20 \gg 1$. This suggests that the normal stress due to the tissue tension $\sigma_{zz}|_{z=h}^{\rm tt}$ is negligible and the boundary condition $\sigma_{zz}\big|_{z=h}=0$ is indeed valid.

%\subsubsection*{Derivation of the full solution}
We may solve Eqs.\ref{P_3d} and \ref{ur_3d}
by applying the method of separation of variable supposing $(1/u_r) \partial^2 u_r / \partial z^2 = k^2 $
or 
$(1/P) \partial^2 P / \partial z^2 = k^2 $
with an arbitrary constant $k$. The general 
solutions for the radial displacement field $u_r$ and the rescaled pressure field $f 
\equiv 2 P/ \hat{G}$ to
Eqs.\ref{ur_3d} - \ref{P_3d} are given as 
\begin{equation}
f(r,z)= \int_{0}^{\infty} dk J_0(kr)\left(C_k e^{kz}+D_k e^{-kz}\right)
\label{noslipgf}
\end{equation}
and
\begin{equation}
u_r(r,z)= \int_{0}^{\infty} dk J_1(kr)\left[ \left( A_k -\frac{C_k z}{2}\right) e^{kz} + 
\left( B_k +\frac{D_k z}{2}\right) e^{-kz}\right]
\label{noslipg}
\end{equation}
with the four coefficients $A_k$, $B_k $, $C_k$ and $D_k$ to be determined with the boundary conditions 
at $z=0$ and $h$
(Eq.\ref{eq:boundary}, main text) and Eq.\ref{uz_3d}. Here, $J_n$ is the Bessel function of first kind of order $n$.
To derive Eqs.
\ref{noslipgf} and \ref{noslipg}, we have used the boundary conditions that
$u_r(r\rightarrow 0)$, 
$u_r(r\rightarrow \infty)$,
$P(r\rightarrow 0)$ and
$P(r\rightarrow \infty)$
are finite.

After a few manipulation,
we obtain explicit forms of $A_k$, $B_k $, $C_k$ and $D_k$
and arrive at the solution
\begin{equation}
 \begin{aligned}
        u_r(r,z,t)= \int_0^\infty dk J_1(kr) \bigg( Q(k,z) \mathcal{H}_1\left\{\frac{S(r,t)}{\hat{G}}\right\} 
        - R(k,z) \mathcal{H}_1\left\{\frac{\xi}{\hat{G}}\frac{\partial u_r}{\partial t}\bigg|_{z=0}\right\}\bigg)
    \end{aligned}
\label{solslip}
\end{equation}
with 
\begin{align}
Q(k,z)=\frac{\mathrm{cosh}(kh)\mathrm{cosh}(kz) - (kh) \mathrm{sinh}(kh)\mathrm{cosh}(kz) + (kz) \mathrm{cosh}(kh)\mathrm{sinh}(kz)}{\mathrm{sinh}(kh) \mathrm{cosh}(kh) + kh}
\end{align}
and
\begin{equation}
    \begin{aligned}
    R(k,z)&=\frac{1}{\mathrm{cosh}(kh) + kh}\Bigg(\mathrm{cosh}(kh)\mathrm{cosh}(k(h-z))
 - (kh) \mathrm{sinh}(kz)\\
 &- (kz) \mathrm{cosh}(kh)\mathrm{sinh}(k(h-z))
+ (kh)(k(h-z)) \mathrm{cosh}(kz)\Bigg)
\end{aligned}
\end{equation}
The notation $\mathcal{H}_n$ means the Hankel transform of order $n$.

Transforming $u_r(r,z,t)$ by Hankel transform of order 1, we obtain $u_r(k,z,t)$ in wavenumber $k-$domain as:
\begin{equation}
    u_r(k,z,t)= \frac{1}{k} Q(k,z) \tilde{S}(k,t)    
    -\frac{\tilde{\xi}}{k} R(k,z)  \frac{\partial u_r(k,0,t)}{\partial t}
\label{uk_3d}
\end{equation}
where $\tilde{S}(k,t)=S(k,t)/\hat{G}$  and $\tilde{\xi}=\xi/\hat{G}$.
For $z=0$, the solution to Eq.\ref{uk_3d} is derived as
\begin{equation}
u_r(k,0,t)=\frac{Q(k,0)}{R(k,0)} \frac{\tilde{S}(k,t)}{\tilde{\xi}}
*M_0(k,t),\label{u0}
\end{equation}
where $M_0$ is the memory kernel 
\begin{equation}
\begin{aligned}
    M_0(k,t)=&\exp\left( -\frac{k}{\tilde{\xi} R(k,0)} t \right) \Theta(t) \\
    =&\exp\left( -\frac{k}{\tilde{\xi}} \frac{\mathrm{sinh}(kh) \mathrm{cosh}(kh) + kh}{\mathrm{cosh}^2(kh) + (kh)^2} t \right) \Theta(t)  .
\end{aligned}
    \label{M0}
\end{equation}
Here, $\ast$ means the 1D convolution over time.
Substituting $u_r(z=0)$ given in Eq.\ref{u0} into Eq.\ref{uk_3d}, 
the final solution is
\begin{equation}
u_r(k,z,t)=S(k,t)*\bigg(m_s(k,z)\delta(t)+m_f(k,z)M_0(k,t)\bigg), 
\label{u_r_memoryform}
\end{equation}
where 
\begin{equation}
\begin{aligned}
    m_s(k,z)=&\frac{1}{\hat{G}}
    \frac{Q(k,z) R(k,0) - Q(k,0) R(k,z)}{k R(k,0)} \\
    =& \frac{1}{\hat{G}k\left(\mathrm{cosh}^2(kh)+(kh)^2 \right)}\Bigg(     \mathrm{cosh}(kh) \mathrm{sinh}(kz) 
    - (kh) \mathrm{sinh}(kh) \mathrm{sinh}(kz)\\
    & +(kz) \mathrm{cosh}(kh) \mathrm{cosh}(kz) 
    - (kh)(kz)\mathrm{sinh}(k(h-z))\Bigg) 
\end{aligned}
\label{eq:fullm_s}
\end{equation}
which is independent of friction, and 
\begin{equation}   
\begin{aligned}
    m_f(k,z)=&\frac{1}{\xi}\frac{Q(k,0) R(k,z)}{R(k,0)^2} \\
    =&\frac{1}{\xi} \frac{\mathrm{cosh}(kh)-(kh)\mathrm{sinh}(kh)}{[ \mathrm{cosh}^2(kh)+(kh)^2 ]^2}  \Bigg ( \mathrm{cosh}(kh)\mathrm{cosh}(k(h-z)) \\
 - &(kh) \mathrm{sinh}(kz)
 - (kz) \mathrm{cosh}(kh)\mathrm{sinh}(k(h-z)) + (kh)(k(h-z)) \mathrm{cosh}(kz) 
\Bigg)
\end{aligned}    
\label{eq:fullm_f}
\end{equation}
which depends on friction.

%General solution
Finally, the general solution can be represented as a space(2D)-time convolution between the surface stress term and a memory kernel as:
\begin{equation}
    \label{eq:generalSolution_3d}
    u_r(r,z,t)=\frac{1}{2\pi}S(r,t)***M(r,z,t),
\end{equation}
where the kernel is
\begin{equation}
    \label{eq:MemoryKernel}
        M(r,z,t)=\mathcal{H}^{-1}_1\{m_s(k,z)\delta(t)+m_f(k,z)M_0(k,t)\},
\end{equation} and the operator ``$***$" represents the convolution of two functions over 2D space and time:
\begin{equation*}
    f_1***f_2(r,t)=\int_{-\infty}^\infty dt' \int_{-\infty}^\infty dr' \int_0^{2\pi} d\phi r' f_1(\vec{r'},t')f_2(\vec{r}-\vec{r'},t-t') .
\end{equation*}
For $r>0$, 
this is equivalent to
\begin{equation}
\begin{aligned}
    f_1***f_2(r,t)=
    2\pi\int_{-\infty}^\infty dt'\int_0^\infty  kdk  J_n(kr) F_{1,n}(k,t')F_{2,n}(k,t-t') \ ,
\end{aligned}
\label{eq:circularconvolution}
\end{equation}
where $J_1$ is the Bessel function of first kind, and $F_{1,n}$ and $F_{2,n}$ are Hankel transforms of $f_1$ and $f_2$ from $r$-domain to the wave-number $k$-domain.
The order $n=1$ is chosen for our solutions.
We also define the notation ``**" for the 2D convolution over space as
\begin{equation}
    f_1**f_2(r)=\int_{-\infty}^\infty dr' \int_0^{2\pi} r' d\phi  f_1(\vec{r'})f_2(\vec{r}-\vec{r'}) ,
\label{eq:2dcircularconvolution}
\end{equation}
with the same convolution identity applied in Eq.\ref{eq:circularconvolution}. 

The dependency of coefficients $m_s$ and $m_f$ on $z$ and $r$ is shown in Fig. \ref{fig:theorySimple}A. The $m_s$ is dominant at the surface ($z=h$) while the friction-dependent kernel $m_f$ plays a stronger role at the bottom ($z=0$). Meanwhile, $m_s$ overwhelms $m_f$ near the center $r<h$, indicating that friction-dependent dynamics is negligible near the center.
%%%%%%%%%%%%%

At the surface $z=h$, Eqs \ref{eq:fullm_s} and \ref{eq:fullm_f} result in
\begin{equation}
\begin{aligned}
    m_s(k,h)=& \frac{1}{\hat{G} k} \frac{\mathrm{cosh}(kh) \mathrm{sinh}(kh) + kh}{\mathrm{cosh}^2(kh)+(kh)^2} 
\end{aligned}
\end{equation}
and 
\begin{equation}
\begin{aligned}
    m_f(k,h)=&\frac{1}{\xi} \Bigg( \frac{\mathrm{cosh}(kh)-(kh)\mathrm{sinh}(kh)}{ \mathrm{cosh}^2(kh)+(kh)^2 } \Bigg)^2 ,
\end{aligned}    
\end{equation}
respectively. Moreover,
$z$-averages of Eqs \ref{eq:fullm_s} and \ref{eq:fullm_f} from $z=0$ to $h$,\\ $\overline{m_{s/f}}\equiv (1/h)\int_0^h dz m_{s/f}(z)$,  are given by 
\begin{equation}
\begin{aligned}
    \overline{m_s}(k)=& \frac{h}{\hat{G}} \frac{1}{\mathrm{cosh}^2(kh)+(kh)^2} 
\end{aligned}
\end{equation}
and 
\begin{equation}
\begin{aligned}
    \overline{m_f}(k)=&\frac{1}{\xi} \frac{\mathrm{cosh}^2(kh)-(kh)\mathrm{sinh}(kh)\mathrm{cosh}(kh)}{ [\mathrm{cosh}^2(kh)+(kh)^2]^2 } ,
\end{aligned}    
\end{equation}
respectively.

As shown in Fig. \ref{fig:theorySimple}A, the coefficient $m_s$ grows with $z$ from zero, and reaches a maximal value at the surface.

When $t>t_c$, the friction dependent part of the kernel $m_f(k,z)M_0(k,t)$ dominates the dynamics with trivial dependency on $z$ and a dynamical scaling emerges from this kernel(see section S3.1). 

\subsection*{S2  Solution under specific surface stress form}
\label{SM:SpecificSolution}
\subsubsection*{S2.1  Solution under a persistent surface stress}
With a persistent surface stress $S(r,t)=S(r)\Theta(t)$, the solution becomes
\begin{equation}
\begin{aligned}
    u_r(r,z,t)
    &=\int_0^\infty kdk J_1(kr) S(k)\left(m_s + m_f\frac{\tilde{\xi}R(k,0)}{k}(1-M_0)\right)\Theta(t)\\
    &=\frac{1}{2\pi}S(r)**M_p(r,z,t),
\end{aligned}
    \label{eq:Persistent}
\end{equation}
where $\mathcal{H}_1\{M_p(r,z,t)\}$ is
\begin{equation}
\begin{aligned}
     M_p(k,z,t)=
     \frac{\Theta(t)}{\hat{G}}\left(\frac{Q(k,z)}{k}- \frac{Q(k,0)R(k,z)}{kR(k,0)}\mathrm{exp}\left[-\frac{k}{\tilde{\xi}R(k,0)}t\right]\right) .
     \label{eq:sol_ps}
\end{aligned}
\end{equation}
The notation $\mathcal{H}_n$ means the Hankel transform of order $n$.

Figure \ref{fig:theorySimple}B shows the $z$-dependency of $u_r$ (normalized by $s\varepsilon/h\hat{G}$ for a specific persistent surface stress $S(r,t)=\Theta(t) s\varepsilon/r$. The far field solution ($r>h$) does not vary too much with the increase of $z$; yet the near field solution ($r<h$) increases with $z$. 

When $t\gg t_c$, or equivalently, when under the 2D approximation $h \ll \hat{G} t / \xi$, this memory kernel $M_p$ is simplified to
\begin{equation}
  M_{p,2D}(r,t) \equiv \frac{\Theta(t)}{\hat{G}}\mathcal{H}^{-1}_1\left\{\frac{1-e^{-2k^2t/\tilde{\tau}}}{2hk^2}\right\},
    \label{eq:M_2D}
\end{equation}
where $\tilde{\tau}=\xi/\hat{G} h$. 

On the contrary, the non-slidable solution ($\xi\to \infty$) is mathematically equivalent to the transient solution at $t\to 0$ :
\begin{equation}
    u_r^n(r,z,t)=\frac{1}{2\pi}S(r)**M_p^n(r,z),
\end{equation}
where $\mathcal{H}_1\{M_n(r,z)\}$ is
\begin{equation}
    M_p^n(k,z)=M_p(k,z,0)=m_s(k,z).
\end{equation}

Therefore, the surface deformation caused by the shear mode when $t\ll t_c$ is
\begin{equation}
    u^n_r(r,h)=\int^\infty_0 dkJ_1(kr) \tilde{S}(k) \frac{\mathrm{sinh}(kh)\mathrm{cosh}(kh)+kh}{\mathrm{cosh}^2(kh)+(kh)^2}.
\end{equation}

This solution under non-slidable limit is none other than the solution in the regime under the pure shearing $u_{\mathrm{shear}}$.
By contrast, the solution under purely stretching regime $u_{\mathrm{stretch}}$ is derived  from the kernel Eq.\ref{eq:M_2D} for the limit $t\to \infty$ or $\xi \to 0$.

To discuss the difference of these two solutions in $r$-dependency, we could integrate them over $z$ axis and it is easily found in mathematics that 
\begin{equation}
\bar{u}_{\mathrm{shear}}(r) =\frac{1}{h}\int_0^h u_r(r,z,\xi \to \infty) dz \propto S(r),
\end{equation}
and 
\begin{equation}
        \ \bar{u}_{\mathrm{stretch}}(r)=\frac{1}{h}\int_0^h u_r(r,z,\xi\to 0)dz \propto \int_0^r S(r')r'dr',
    \label{eq:sol_zero_fric}
\end{equation}
respectively. Clearly, $u_{\mathrm{stretch}}$ is the accumulated local deformations from the center ($r=0$), which has a distinct radial dependency from that of  $u_{\mathrm{shear}}$. Undedr the assumption that the surface stress decaying over space, $S(r) \sim 1/r^q$, $u_r$ in the shear regime has a negative $r$-dependence while $u_r$ in the stretch regime accumulates with $r$ from the center, $r=0$ (referring to the small $t$ and large $t$ region in Fig. \ref{fig:disc}C in the main text); hence, a peak emerges at the boundary of two regimes.

\subsubsection*{S2.2  Solution under persistent surface stress with a cut-off length}
In our experiments, the peaks of gel displacement propagate over a distance of 2-3 cell diameters from the center in experiments. Our theory suggests that the propagation distance could be constrained by a cutoff range around the apoptotic cell. Mechanical responses can have long-range correlations in tissues \cite{fletcher2010cell,rossen2014long,charras2018tensile} and the correlation length scale is determined by the competition between strength of intercellular mechanics and other effects that decorrelate the forces among cells. Therefore, the cutoff range in our theory is the effective area in which the intercellular force transmissions overpowers the decorrelating factors such as the intrinsic cell activities and thermal noises from subcellular and extracellular environment. This range can be associated with stiffness, fluidity (cell-cell rearrangement) and cell cycle activities of a tissue.

We add the ingredient of the cutoff length $l$ to the surface stress by an exponential decaying term:
\begin{equation}
  S(r,t)=s \left( \frac{\varepsilon}{r} \right)^{q} e^{-r/l}\Theta(t), 
    \label{eq:spreadingWithRange}
\end{equation}
and as shown in \ref{fig:nondif}B, the propagation stops around $r^\infty_{peak}\sim 1.5 l$ for $q=1$. This could be understood from the following calculations.

Substituting the form Eq.\ref{eq:spreadingWithRange} in to Eq.\ref{eq:Persistent} gives the solution as:
\begin{equation}
 u_r(r,z,t)=
    s\varepsilon\int_0^\infty kdk J_1(kr)\frac{k}{K(K+k)}M_p(k,z,t),
    \label{solutionWithDecay}
\end{equation} 
where $K=\sqrt{k^2+1/l^2}$. The kernel $k/\left(K(K+k)\right)$ is exactly the Hankel transform of $e^{-r/l}/r$. By rewriting Eq.\ref{solutionWithDecay} into a convolution form we get:
\begin{equation}
    u_r(r,z,t)=\frac{s\varepsilon}{2\pi}\frac{1}{r}**\mathcal{H}^{-1}_1\{M_l(k,z,t)\},
\end{equation}
where 
\begin{equation}
  M_l(k,z,t)= \frac{\Theta(t)}{\hat{G}}\frac{1}{h K(K+k)} M_p(k,t)
\end{equation}

Since we would like to know long-term behavior of the model, we can use the 2D approximation ($\xi h/\hat{G}\ll t$) and trivialize the $z$-dependency so that
\begin{equation}
     M_{l,2D}(k,t) \equiv \frac{\Theta(t)}{\hat{G}}\frac{1-e^{-2k^2t/\tilde{\tau}}}{2hK(K+k)},  
\end{equation}
where $\tilde{\tau}=\xi/\hat{G}h$.
Clearly, when $t/\tilde{\tau}\ll l^2$, $M_{l, 2D}(r,t)$ could be reduced to 
$M_{p,2D}$ (Eq.\ref{eq:M_2D}), in which the diffusive scaling is well preserved. By contrast, for long time limit 
$t/\tilde{\tau} \gg l^2$,
the $M_{l, 2D}(r,t)$ is reduced to a static form that spatially scales with $l$ as:  
\begin{equation}
\begin{aligned}
    M_{l,2D}(r,t\gg l^2\tilde{\tau})&=\widetilde{M}_{l,2D}(\tilde{r}) \text{ with } \tilde{r} \equiv r/l \\
    &=\frac{\Theta(t)}{2\hat{G}}\mathcal{H}^{-1}_1\left\{\frac{1}{h\sqrt{\tilde{k}^2+1}(\sqrt{\tilde{k}^2+1}+\tilde{k})}\right\} ,
\end{aligned}
\end{equation}
where the inverse Hankel transform ${H}^{-1}_1$ is applied from the normalized wavenumber domain $\tilde{k}=kl$ to the normalized spatial domain $\tilde{r}=r/l$.

\subsubsection*{S2.3  Solution with a decaying surface stress}
Another aspect to consider is a decaying surface stress with time.
This could happen due to multiple reasons such as the viscoelastic relaxations of tensions in the cytoskeleton as well as in connective integrins and the formation of contractile actin-myosin rings around the extruding cell.
Fig. \ref{fig:nondif}B shows the solution with a linear decaying stress:
\begin{equation}
    S(r,t)=s \left( \frac{\varepsilon}{r} \right)^{q} \Theta(t)(1-ct), 
    \label{eq:dissipativeSource}
\end{equation}
where $c\ll 1/t_c$ quantifies a slow decay with time for $q=1$. The displacement $u_r$ near the center turns to decrease over longer time (top panel of Fig. \ref{fig:nondif}C), indicating that the gel near the center is gradually moving inward to the center and meanwhile the magnitude of peak $u_{peak}$ exhibits a nonmonotonic trend of evolution (Fig. \ref{fig:nondif}B, bottom). These results can qualitatively explain the emergence of a inward movement region near the center in the experiment in the main text(Fig. \ref{fig:scheme}B, blue region) and the nonmonotonicity of the  $u_{peak}$ dynamics (Fig. \ref{fig:peak}A, bottom). 

The detailed calculation of the memory kernel under 2D approximation is
\begin{equation}
    \begin{aligned}
        M_{d}(k,t)&=\frac{\Theta(t)}{2\hat{G}}\int_0^{t} \frac{e^{-2k^2(t-t')/\tilde{\tau}}}{h\tilde{\tau}}(1-ct')dt'\\
        &= M_{2D}(k,t) - c M_{c}(k,t)
    \end{aligned}
    \label{Td}
\end{equation} 
with
\begin{equation}
        M_{c}(k,t) = \frac{\Theta(t)}{2\hat{G}}\frac{\tilde{\tau}}{h}\frac{2k^2t/\tilde{\tau}+e^{-2k^2t/\tilde{\tau}}-1}{4k^4} , \notag 
\end{equation}
and the final solution $u_r(r,t)$ is
\begin{equation}
    u_r(r,t)= \bar{u}_{2D}-\frac{c}{2\pi}S(r)**\mathcal{H}^{-1}_1\{M_{c} (k,t) \}.
\end{equation}
The inverse Hankel transform of $M_c(k,t)$ has the profile similar to that of $M_{2D}$ except that the magnitude of $M_c(k,t)$ grows with time $t^2/\tilde{\tau}$ for $r<\sqrt{2\hat{G}ht/\xi}$. Hence, the final profile of $u_r(r,t)$ has a negative velocity for $r<r_{peak}$.

%One possible mechanism for this slow decay is the natural relaxation of forces among cells or between cell and substrate as the substrate deforms \cite{Khalilgharibi2019}. Tissue viscoelasticity may also play a role to induce temporal changes in the spatial stress distributions. Another possible mechanism could be the appearance of forces globally pulling the cells back into the center. The timescale for E-cadherins to accumulate around the extruding cell  is roughly 20 minutes after the ablation \cite{lubkov2014cadherin,takeuchi2020calcium}, comparable to the onset time of stage II in experiments and therefore, the formation of contractile actin-myosin ring around the extruding cell could be another viable mechanism for the slow decay.  

\subsection*{S3  Calculating integrals with Bessel functions}
\label{SM:integral}
In this and the following sections, we provide the detail calculations on the most relevant solutions in a weakly adhesive (slidable) case and in a rigidly adhesive (non-slidable) case. In the calculations, integrals with Bessel functions are the key to the solutions and the technique of nondimensionalizing the variables is extensively used for deriving both rigorous and asymptotic solutions. 

\subsubsection*{S3.1  Dynamic scaling of the memory kernel $M(r,z,t)$}
$M(r,z,t)$ can be decomposed into two parts, a source- dependent part
\begin{equation*}
    M_s(r,z,t)=\mathcal{H}^{-1}_1\{m_s(k,z)\delta(t)\}
\end{equation*}
and a friction dependent part
\begin{equation*}
    M_f(r,z,t)=\mathcal{H}^{-1}_1\{m_f(k,z)M_0(k,t)\}.
\end{equation*}

The nature of diffusive propagation originates from the dynamic scaling of the friction dependent memory kernel $M_f(r,z,t)$ when the elapsed time $t$ surpasses a critical scale $t_c=h\xi/\hat{G}$, i.e., the gel's thickness $h$ is relatively small as compared with a critical length $t\hat{G}/\xi$ and $z$ dependency becomes trivial. This is seen as follows:

The original friction-dependent kernel in $k$ domain is:  
\begin{equation}
    M_f(k,z,t)=m_f(k,z)e^{-tk\hat{G}/(\xi R(k,0))},
\end{equation}
where $R(k,0)$ in full model is 
\begin{equation}
    R(k,0)=\frac{\mathrm{cosh}^2(kh)+(kh)^2}{\mathrm{sinh}(kh)\mathrm{cosh}(kh)+kh}.
\end{equation}
Substituting $\tilde{k}=kt\hat{G}/\xi$ into $M_f$, we then get:
\begin{equation}
    M_f(\tilde{k},z,t)=m_f(\tilde{k},z)e^{-\tilde{k}/R(\tilde{k}\xi/\hat{G}t,0)}.
\end{equation}
When $t\gg t_c=h\xi/\hat{G}$, $\mathrm{sinh}(kh)\to kh$ and $\mathrm{cosh}(kh)\to 1$, thus 
the kernel keeps its time-dependency in the exponential terms in such a way:

\begin{equation}
    M_f(\tilde{k},z,t)=m_f(\tilde{k},z)e^{-2\tilde{k}^2 \xi h /\hat{G}t}
    \label{eq:reducedfriction-dependentkernel}
\end{equation}
and consequently in $r-$domain:
\begin{equation}
    M_f(r,z,t)=\int_0^\infty  m_f(\tilde{k},z)e^{-2\tilde{k}^2\xi h /\hat{G}t}J_1(kr)kdk.
    \label{eq:M_f-rdomain}
\end{equation}
Note that $m_f(\tilde{k},z)=m_f(k,z,h\to 0)\sim 1/\xi $ when $t\gg t_c$ because the $z$-dependence becomes trivial.

Let $k' = \tilde{k} \sqrt{2\xi h/\hat{G}t}=k\sqrt{2\hat{G}th/\xi}$ and substitute into Eq.\ref{eq:M_f-rdomain},
we finally get
\begin{equation}
    M_f(r,z,t)= \frac{\xi}{2\hat{G}ht}\int_0^\infty m_f(k',z) e^{-k'^2}J_1(k'r/r_c)k'dk', 
\end{equation}
and thus\begin{equation}
    M_f(r,z,t)=\frac{\xi}{2\hat{G}ht}\widetilde{M}_f\left(\frac{r}{r_c},z \right),\label{eq:Mf_scaling}
\end{equation}
where $r_c=\sqrt{2\hat{G}ht/\xi}$ and $\widetilde{M}_f$ is the rescaled form 
\begin{equation}
    \widetilde{M}_f=\mathcal{H}^{-1}_1\{m_f(k,z,h\to 0)e^{-k^2}\},
\end{equation}
where the notation $\mathcal{H}^{-1}_1$ means the Inverse Hankel transform of order 1. This spatial scaling  $r_c \propto \sqrt{t}$ in $M_f$ is the origin of a diffusive propagation in the final solution and if the surface stress field $S(r,t)$ decays in space as $S(r)\propto (1/r)^q\ (0<q<2)$, $r_c$ is exactly where $r_{peak}$ locates. This explains why the peak starts to propagate at $r\sim h$ ($r_c$ only exists when $t>\xi/\hat{G}$, hence $r_c> h$). 

\subsubsection*{S3.2  Peak magnitude in 2D approximation}
As explained in the main text, when the elapsed time surpasses a critical scale $t_c=\xi h/\hat{G}$, the role of thickness of the gel is trivial and thus the solution for the $z-$averaged displacement $\bar{u}$ can be representative for the 3D dynamics. With the surface stress $S(r,t)=s(\varepsilon/r)^q\Theta(r)\Theta(t)$, this solution is 
\begin{equation}
    \bar{u}_r= \frac{s}{2\pi}\left(\frac{\varepsilon}{r}\right)^q**M_{2D}(r,t),
    \label{eq:originalConvolution}
\end{equation}
where "**" is the 2D convolution over space, and the memory kernel $M_{2D}$ is
\begin{equation}
    M_{2D}(r,t)=\mathcal{H}^{-1}\left\{\frac{1-e^{-2k^2t\hat{G}h/\xi}}{2k^2h\hat{G}}\right\}.
\end{equation}
For sake of  simplicity, $\Theta(t)$ is dropped in all the following calculations.
The solution under 2D approximation ($h<\hat{G} t/\xi$) has a pronounced propagating peak, and the slope rising to the peak and the slope decaying from the peak can be rigorously obtained by setting $t\to 0$ and $t\to \infty$. With $t\to0$, $M_{2D}$ is reduced to $\mathcal{H}^{-1}\{t/\xi\}$, then the solution is 
\begin{equation}
    \bar{u}_r(r,t\to 0)=\frac{st}{\xi} \left(\frac{\varepsilon}{r}\right)^q,
\end{equation}
showing a decaying power the same as in the surface stress $S(r)$.

With $t\to \infty$, $M_{2D}$ is reduced to $\mathcal{H}^{-1}\{1/k^2\}/(2h\hat{G})$, then the solution is
\begin{equation}
    \bar{u}_r(r,t\to \infty)\propto r^{2-q},
\end{equation}
showing a rising power $2-q$ if $0<q<2$. 

The displacement field crosses over from $\bar{u}(r,t\to 0)$  to $\bar{u}(r,t\to \infty)$ in space and the turning point occurs at somewhere $r_c\propto \sqrt{t}$. Due to the nature of crossover, the magnitude of peak $u_{peak}(r=r_c)$ is not directly the intersection point of $\bar{u}_r(r,t\to 0)$ and $\bar{u}_r(r,t\to \infty)$. To calculate the peak magnitude, we need a rigorous inspection on the convolution Eq.\ref{eq:originalConvolution}. As a simple demonstration, we calculate the case with $q=1$ as follows. 

The 2D convolution form can be represented as an integral with a Bessel function:
\begin{equation}
    \bar{u}_r(r,t)=\frac{s\varepsilon}{\hat{G}}\int_0^\infty dk J_1(kr) \frac{1-e^{-2k^2t\hat{G}h/\xi}}{2k^2h}.
    \label{eq:Besselintegral}
\end{equation}
By substituting $k'=k\sqrt{2t\hat{G}h/\xi}=kr_c$, which is a dimensionless wavenumber, into Eq.\ref{eq:Besselintegral}, we get
\begin{equation}
     \bar{u}_r(r,t) = \frac{s\varepsilon r_c}{2\hat{G}h} \int_0^\infty dk'J_1\left(k'r'\right)\frac{1-e^{-{k'}^2}}{{k'}^2} ,
     \label{eq:dimensionless}
\end{equation}
where $r'= r/r_c$ is a dimensionless radial distance. It is easily found that the integral in Eq.\ref{eq:dimensionless} peaks at $\tilde{r}=1$ and the value of the peak is a constant
\begin{equation}
    Z=\int_0^\infty dxJ_1(x)\frac{1-e^{-x^2}}{x^2}\sim 0.48227,
\end{equation}
and finally the peak magnitude $u_{peak}$ is found as
\begin{equation}
    u_{peak}= Z\frac{s\varepsilon r_c}{2\hat{G}h}= Z s\varepsilon\sqrt{\frac{t}{2 \hat{G}\xi h}}.
\end{equation}
This skill of nondimensionalizing $k$ in the integral is also valid for the calculations with $q$ other than 1.

\subsubsection*{S3.3  Solution in a non-slidable limit}
In a non-slidable case, the friction $\xi\to \infty$, and the 2D approximation above is invalid. As explained before, the friction-dependent kernel $M_f$ would vanish and then the solution at the surface $z=h$ becomes
\begin{equation}
    u^n_r(r,h,t)=\frac{S(r,t)}{2\pi}**m_s(k,h)=\frac{s\varepsilon}{\hat{G}} \int^\infty_0 dk J_1(kr) \frac{\mathrm{sinh}(kh)\mathrm{cosh}(kh)+kh}{k[\mathrm{cosh}^2(kh)+(kh)^2]}
    \label{eq:ConvNonSlip}
\end{equation}
with the surface stress $S(r,t)=s(\varepsilon/r)\Theta(r)\Theta(t)$.

By substituting $k$ with $\tilde{k}=kh$, we get:
\begin{equation}
    u^n(r,h)=\frac{s\varepsilon}{\hat{G}} \int^\infty_0 d\tilde{k} J_1(\tilde{k} r/h) \frac{\mathrm{sinh}(\tilde{k})\mathrm{cosh}(\tilde{k})+\tilde{k}}{\tilde{k}[\mathrm{cosh}^2(\tilde{k})+\tilde{k}^2]},
    \label{eq:ndSolNonSlip}
\end{equation}
which is $\sim s\epsilon/\hat{G} $ for $r<h$ and decays with $1/r$ for $r>h$ (Fig. \ref{fig:SI}A). 

\subsubsection*{S3.4  Role of spatial onset of surface stress}

Let's consider the surface stress $S(r,t)=S(r)\Theta(r-r_0)\Theta(t)$ ($r_0>0$). Substituting this stress field into the general convolution, we get the solution in 2D approximation for the slidable case as:
\begin{equation}
\bar{u}_r(r,t)= s\varepsilon \int_0^\infty kdk J_1(kr)J_0(kr_0)S(k)M_{2D}(k,t),   
\end{equation}
by comparing which with Eq.\ref{eq:Besselintegral}, we can see the only difference lies in the emergence of the factor $J_0(kr_0)$, which equals 1 when $r_0=0$. 

By nondimensionalizing $k$, we can find that when $r_c>r_0$, i.e., $t>r_0^2\xi/\hat{G}$, the $J_0\to 1$, so that we would get the same result with Eq.\ref{eq:Besselintegral}. Now remember 2D approximation $t>t_c$ is the condition for Eq.\ref{eq:Besselintegral} to be derived, and if the onset position $r_0$ is smaller than the gel thickness $h$ (this is almost the case in all our experiments), the critical timescale (brought by nonzero $r_0$) $r_0^2\xi/\hat{G}$ is smaller than $t_c$. Hence the role of $r_0$ can be neglected in 2D approximation for a slidable case. Since the propagating dynamics in 3D has trivial dependency on the height of gel for $r>h$, any statement made with 2D approximation can hold for the propagation in 3D. In conclusion, the role of $r_0$ is trivial to the propagation of peak.

Similarly, we can understand the role of $r_0$ in a non-slidable limit by substituting the surface stress with $r_0$ into Eq.\ref{eq:ConvNonSlip} and arrive at:
\begin{equation}
    u_r^n(r,h)=\frac{s\varepsilon}{\hat{G}}\int_0^\infty dk J_1(kr)J_0(kr_0)\frac{\mathrm{sinh}(kh)\mathrm{cosh}(kh)+kh}{k[\mathrm{cosh}^2(kh)+(kh)^2]}. \label{eq:nonBsIntegralr0}
\end{equation}

By nondimensionalizing $k$, we can find that for $r>r_0$, the $J_0$ factor is trivial so that the solution Eq.\ref{eq:ndSolNonSlip} still holds. But for range $r<r_0$, the $J_0$ factor plays a role as shown in Fig. \ref{fig:SI}B. The deformation occurs most at where $r>r_0$ and $r<h$. 

\subsection*{S4  Simplified theory and the solution}
\label{SM:simleModel}
We calculated a simplified model in which the term representing the pressure gradient in Eq.\ref{eq:ur_3d} (main text) is neglected. Even without this pressure term, the theory still reproduces the above-mentioned results only with a magnitude difference of $\sqrt{2}$. Therefore, the pressure %inhomogeneity 
gradient induced by the gel incompressibility is not essential for this peak propagation. 

Here we describe the details of the simplified theory. 
Instead of Eq.\ref{ur_3d} we use
\begin{equation}
    \frac{\partial^2 u_r}{\partial r^2} + \frac{1}{r}\frac{\partial u_r}{\partial r} - \frac{u_r}{r^2} + \frac{\partial^2 u_r}{\partial z^2} = 0,
\label{eq:ur_3dSimple}
\end{equation}
 where $\partial P/\partial r$ has been ignored.
The boundary conditions are
\begin{equation}
    \left\{
    \begin{aligned}
        \sigma_{rz} \big|_{z=h}&=S(r,t)\\
        \sigma_{rz} \big|_{z=0}&=\xi\frac{\partial u_r}{\partial t}\bigg|_{z=0}\\
        \frac{\partial u_z}{\partial r}\big|_{z=0} &=0 ,
    \end{aligned}
    \right.
    \label{eq:boundarySimple} 
\end{equation}
where $\sigma_{rz}$ is given by Eq.\ref{sigma_rz}.
To complete the calculation, we also use the incompressible condition (Eq.\ref{eq:incompressibility}).

Through the same method as that used in Section %\ref{sec:fulltheoryderivation} of SI, 
Eq.\ref{eq:ur_3dSimple} leads to 
\begin{equation}
u_r(r,z)= \int_{0}^{\infty} dk J_1(kr)\left(A_k e^{kz}+B_k e^{-kz}\right),  
\label{eq:noslipgSimple}
\end{equation}
with only two integral coefficients $A_k$ and $B_k$. Using the boundary conditions given in Eq.\ref{eq:boundarySimple}, we can determine $A_k$ and $B_k$ and eventually obtain 
Eq.\ref{solslip} with
\begin{equation}
  Q(k,z)= \frac{\mathrm{cosh}(kz)}{\mathrm{sinh}(kh)}
  \label{eq:simple_Q}
\end{equation}
and
\begin{equation}
  R(k,z)= \frac{2\mathrm{cosh}(k(h-z))-\mathrm{cosh}(z)}{\mathrm{sinh}(kh)}
  \label{eq:simple_R}
\end{equation}
as the solution.
Hence, by applying the same manipulations as those performed to derive the full solution in the last Section, 
we obtain
Eq.\ref{u_r_memoryform} with 
the memory kernel
\begin{equation}
    M_0(k,t)=\mathrm{exp}\left[-\frac{1}{\tilde{\xi}}\frac{k \mathrm{sinh}(kh)}{2\mathrm{cosh}(kh)-1}t \right]\Theta(t) 
    \label{eq:M0Simple}, 
\end{equation}
the friction-independent part
\begin{equation}
    m_s(k,z)=\frac{1}{\hat{G}k}\frac{\mathrm{sinh}(kz)}{2 \mathrm{cosh}(kh)-1}    
\end{equation}
and the friction-dependent part
\begin{equation}
    m_f(k,z)=\frac{1}{\xi}\frac{2\mathrm{cosh}(k(h-z))-\mathrm{cosh}(kz)}{[2\mathrm{cosh}(kh)-1]^2} .
\end{equation}
At the bottom $z=0$, these results fall into
\begin{equation}
\begin{aligned}
    m_s(k,0)=& 0
\end{aligned}
\end{equation}
and 
\begin{equation}
\begin{aligned}
    m_f(k,0)=&\frac{1}{\xi} \frac{1}{ 2\mathrm{cosh}(kh)-1 } ,
\end{aligned}    
\end{equation}
respectively. Moreover,
taking $z$-averages from $z=0$ to $h$, $\overline{m_{s/f}}\equiv (1/h)\int_0^h dz m_{s/f}(z)$, gives
\begin{equation}
\begin{aligned}
    \overline{m_s}(k)=& \frac{1}{\hat{G} h} \frac{\mathrm{cosh}(kh)-1}{ k^2  [2\mathrm{cosh}(kh)-1]} 
\end{aligned}
\end{equation}
and 
\begin{equation}
\begin{aligned}
    \overline{m_f}(k)=&\frac{1}{\xi h} \frac{\mathrm{sinh}(kh)}{ k  [2\mathrm{cosh}(kh)-1]^2 } ,
\end{aligned}    
\end{equation}
respectively.

Figure \ref{fig:theorySimple}(C and D) show the results of this simplified theory in the format used for the full theory (Fig. \ref{fig:theory} of the main text), by which we can compare the results of the full theory and this simplified theory. As seen there, all the results of the simple theory (brown curves) differs from the full model (black curves) slightly in the near-center range (small $r$), yet the features of the peak propagation together with the $1/r$ decay in the far field are reproduced by the simplified model.
This means that, for the mechanism giving rise to the peak propagation, the pressure %inhomogeneity 
gradient
and the coupling between $u_r$ and $u_z$ are not important. The propagation of $u_r$ itself is the essence of this single-peak wave propagation.

%%% Each figure should be on its own page

% For your review copy (i.e., the file you initially send in for
% evaluation), you can use the {figure} environment and the
% \includegraphics command to stream your figures into the text, placing
% all figures at the end.  For the final, revised manuscript for
% acceptance and production, however, PostScript or other graphics
% should not be streamed into your compliled file.  Instead, set
% captions as simple paragraphs (with a \noindent tag), setting them
% off from the rest of the text with a \clearpage as shown  below, and
% submit figures as separate files according to the Art Department's
% instructions.

\subsection*{Figs. S1 to S5}
\clearpage
\setcounter{figure}{0}
\renewcommand{\thefigure}{S\arabic{figure}}
\begin{figure}
    \centering
    \includegraphics[width=0.8\linewidth]{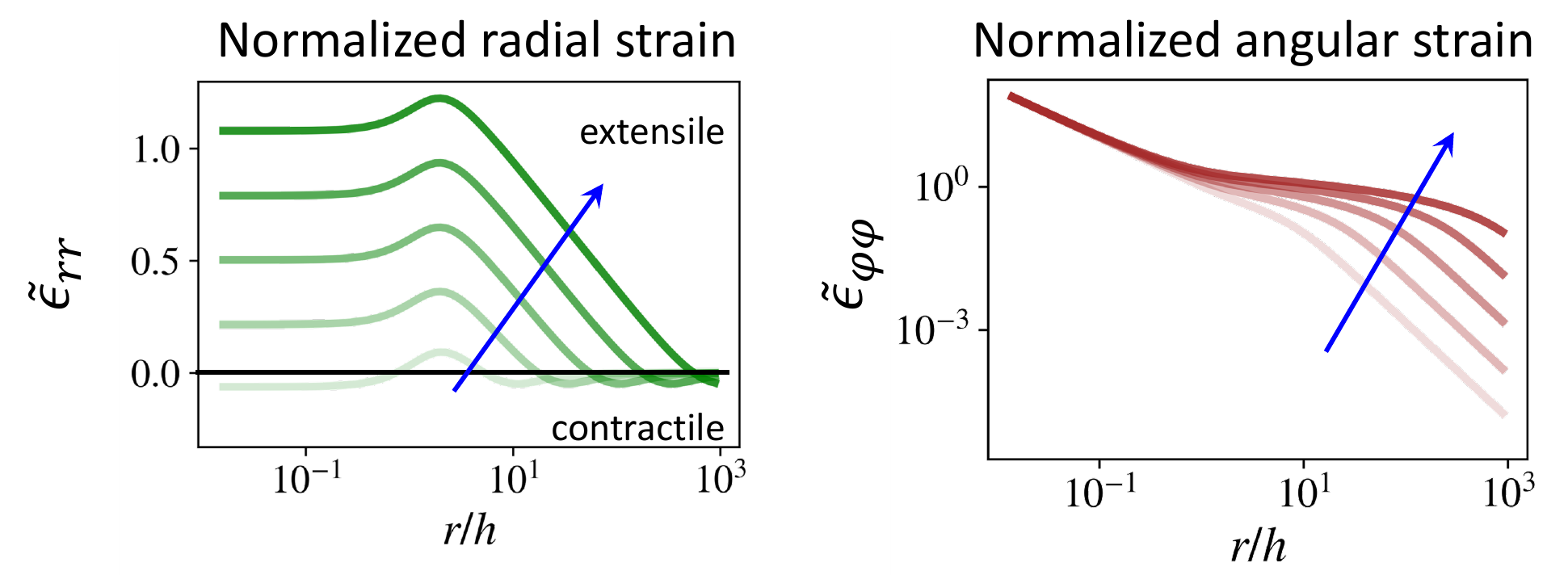}
    \caption{ Numerical results of normalized radial strain $\tilde{\epsilon}_{rr}$ and normalized angular strain $\tilde{\epsilon}_{\varphi\varphi}$ calculated from our model at the substrate surface ($z=h$) under persistent surface stress $S(r,t)=s\varepsilon/r\Theta(t)$. The normalized radial strain $\tilde{\epsilon}_{rr} = \partial \tilde{u}_r(z=h)/\partial (r/h)$ and the normalized angular strain $\tilde{\epsilon}_{\varphi\varphi} = \tilde{u}_r(z=h)h/r$, where $\tilde{u}_r$ is the displacement normalized by $s\varepsilon/h\hat{G}$. 
    Note that the value of radial strain $\epsilon_{rr}$ shifts from positive (extensile strain) to negative (contractile strain) when $r$ surpasses $r_{peak}$, whereas the value of angular strain $\epsilon_{\varphi\varphi}$ keeps positive. Blue arrows denotes the direction of time.}
    \label{fig:strain}
\end{figure}

\begin{figure}
    \centering
    \includegraphics[width=0.8\linewidth]{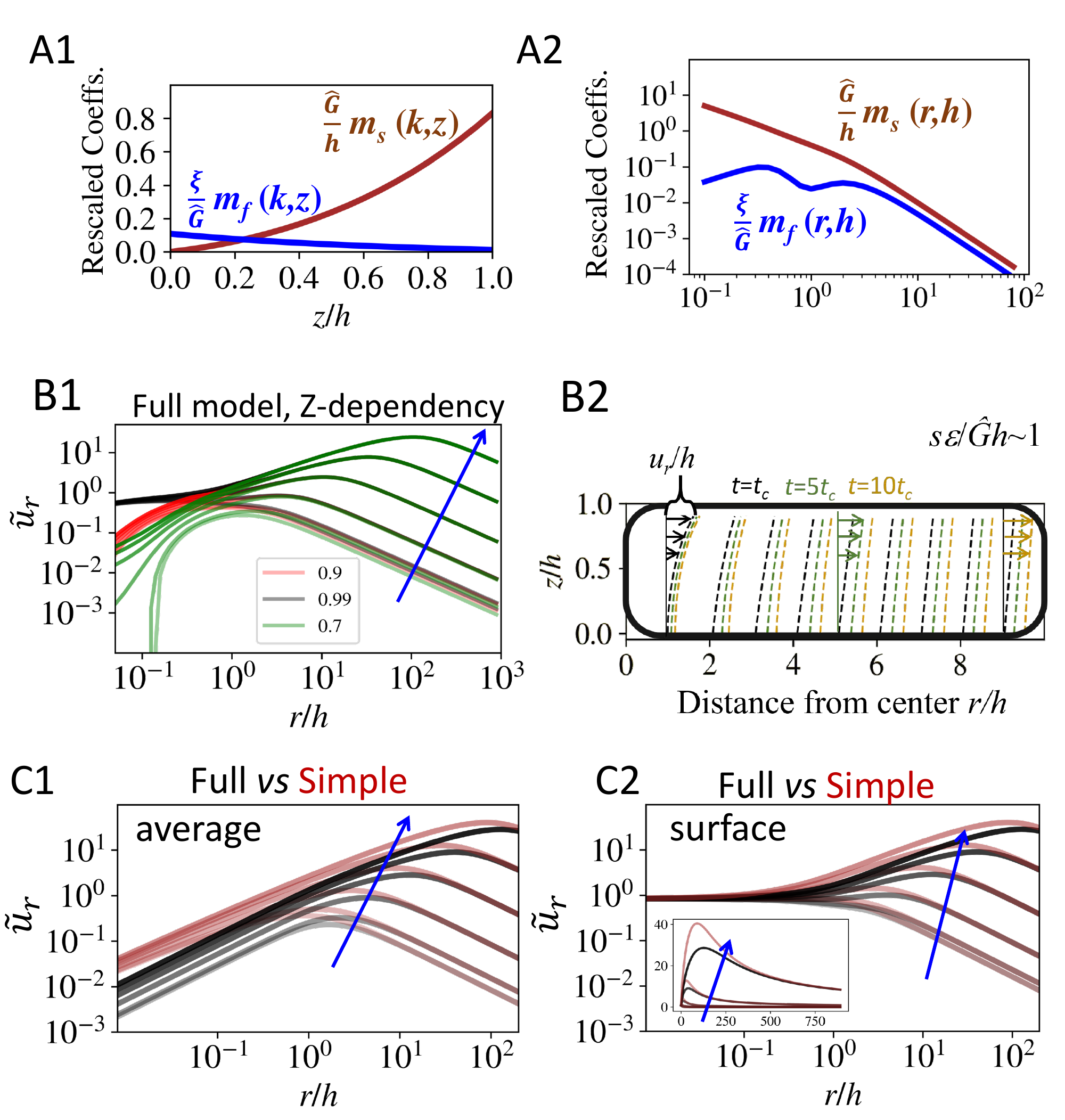}
    \caption{ Model results. (A) Top: The $z$ dependency of two coefficients $m_s$ and $m_f$ in the memory kernel for $k= 1/h$. 
Bottom: the $r$ dependency of $m_s$ and $m_f$ for $z=h$. (B) The radial displacement $u_r$ (normalized by $s\varepsilon/h\hat{G}$) for different time $t$ under the persistent surface stress $S(r,t)=s\varepsilon/\hat{G}\Theta(t)$. (B1) The comparison between different height $z$ with respect to $h$. Blue arrows denote the time evolution. (B2) The displacement plotted in $r-z$ space plane. For a better visibility of the deformation, the parameter is set to $s\varepsilon/\hat{G}h \sim 1$. (C) The comparison between the full model (black curves) and the simple model (brown curves). (C1) Solution averaged over $z$. (C2) Solution at surface $z=h$. The main figure is plotted in the log scale whereas the inset is in the linear scale.}
    \label{fig:theorySimple}
\end{figure}

\begin{figure}
    \centering
    \includegraphics[width=0.8\linewidth]{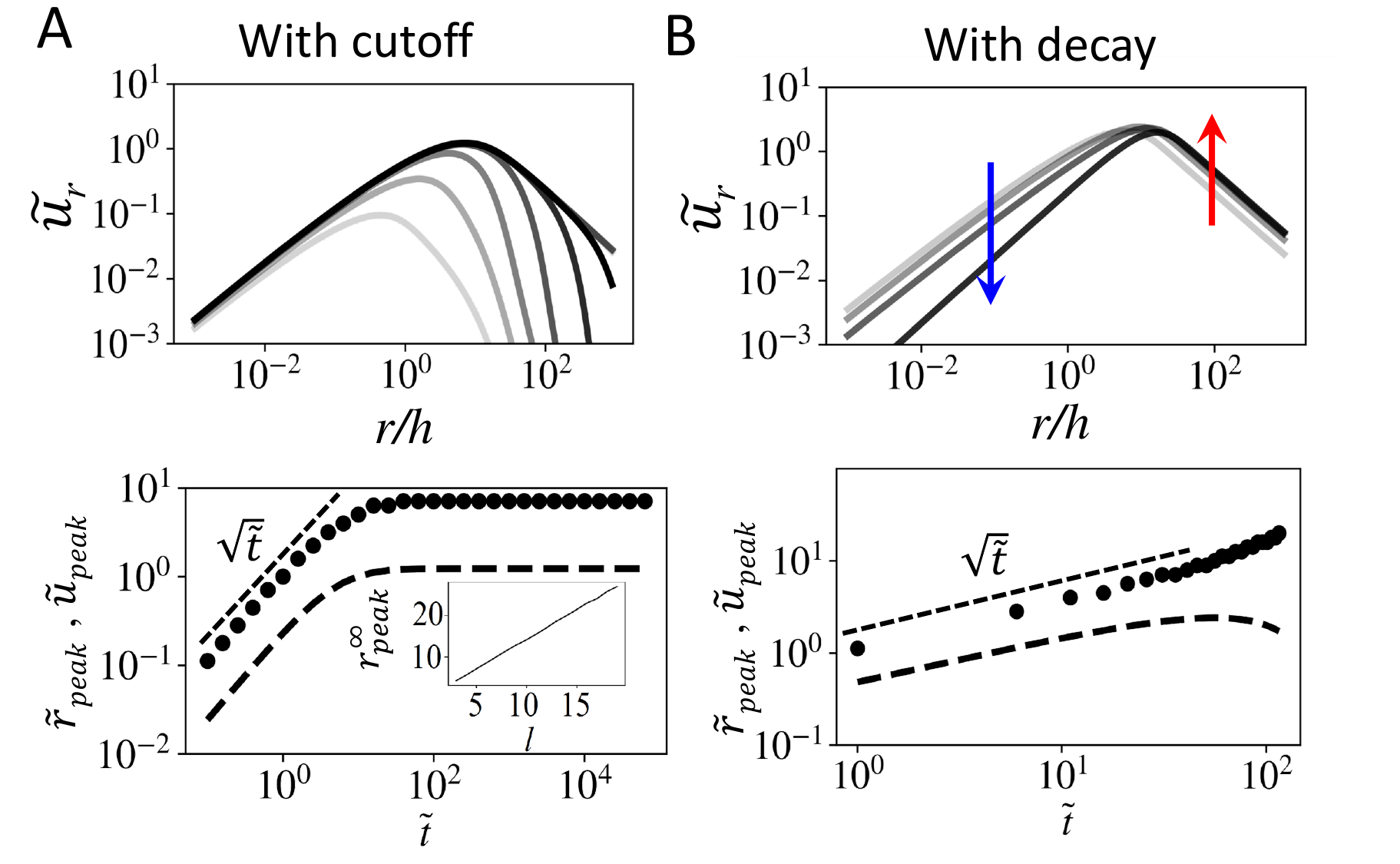}
    \caption{ Longtime non-diffusive propagation dynamics in theory. (A) Top: Solution under the surface stress with a cutoff length $l$ (Eq.\ref{eq:spreadingWithRange}) in the model with 2D approximation. Top: Radial displacement $u_r$ normalized by $s\varepsilon/h\hat{G}$ for varying time, with $l=5h$. Bottom: Dynamics of $\tilde{r}_{peak}$ (dots) and $\tilde{u}_{peak}$ (dashed line) with dimensionless time $\tilde{t}$. Propagation stops at $r_{peak}\sim 1.5l $. Inset: the furthest peak position $r^\infty_{peak}$ is roughly $1.5l$. (B) Solution under the surface stress with a linear decay rate $c$ (Eq.\ref{eq:dissipativeSource}) in the model with 2D approximation. Top: Radial displacement $\tilde{u}_r$ normalized by $s\varepsilon/h\hat{G}$ for varying time for $c=0.01$. Blue arrow indicates the negative velocity of displacement (inward movement) and the red arrow indicates the positive velocity of displacement (outward movement). Right: Dynamics of $\tilde{r}_{peak}$ (dots) and $\tilde{u}_{peak}$ (dashed line) with dimensionless time $\tilde{t}$. A nonmonotonic trend in $\tilde{u}_{peak}$ appears for larger time. }
    \label{fig:nondif}
\end{figure}

\begin{figure}
    \centering
    \includegraphics[width=0.9\linewidth]{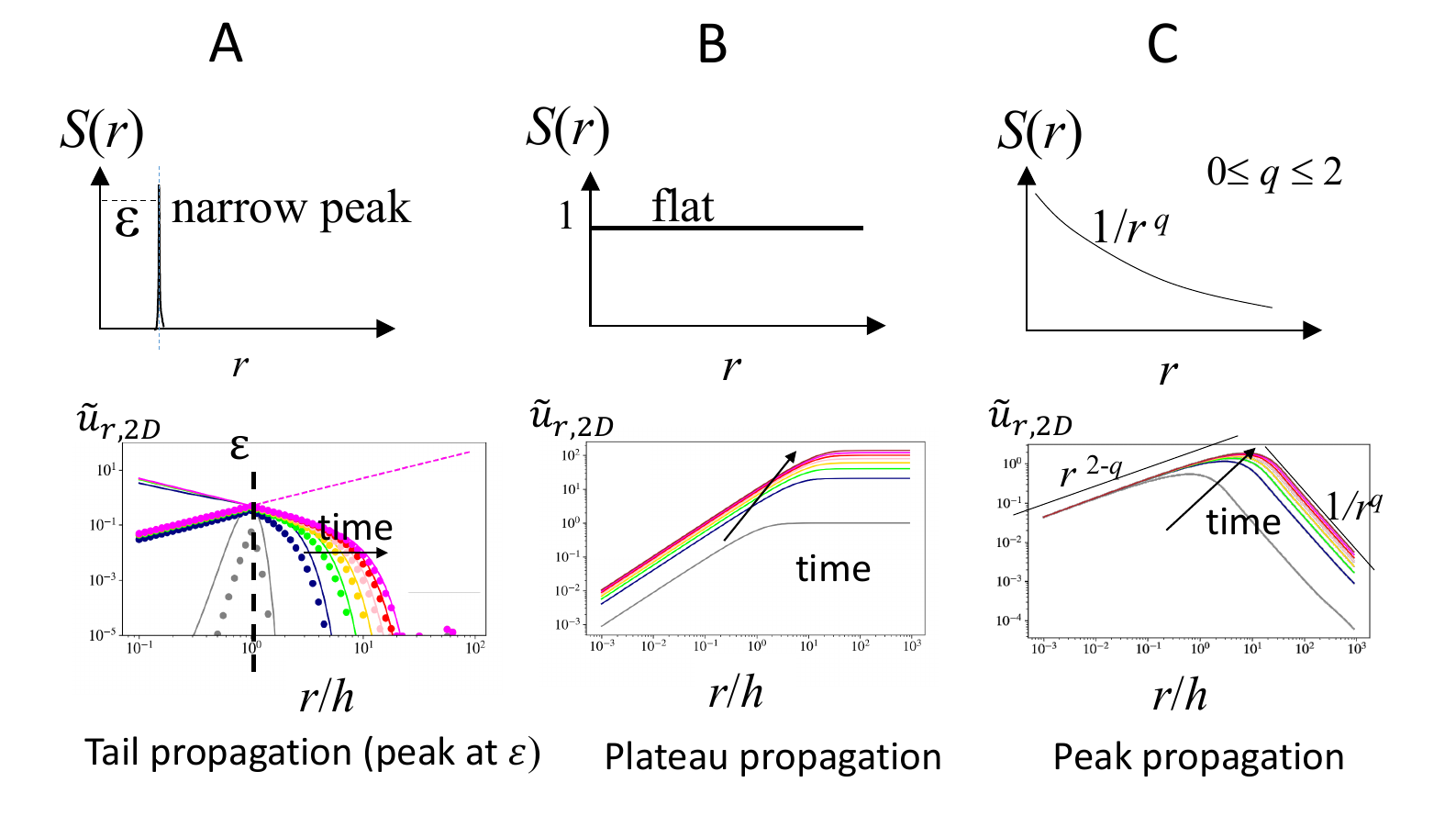}
    \caption{ The radial displacement in 2D approximation $u_{r,2D}(r,t)$ under three different forms of persistent surface stress: (A) a narrow peak $S(r)=\delta(r-\varepsilon)$, where $\delta$ is the Dirac function ; (B) A flat distribution $S(r)=1$; (C) A power-law decay $S(r)=(\varepsilon/r)^q \  (0\le q\le2)$. Peaky profile only appears when $S(r)$ obeys a power-law decay.}
    \label{fig:s-dep}
\end{figure}

\begin{figure}
\centering
\includegraphics[width=\textwidth]{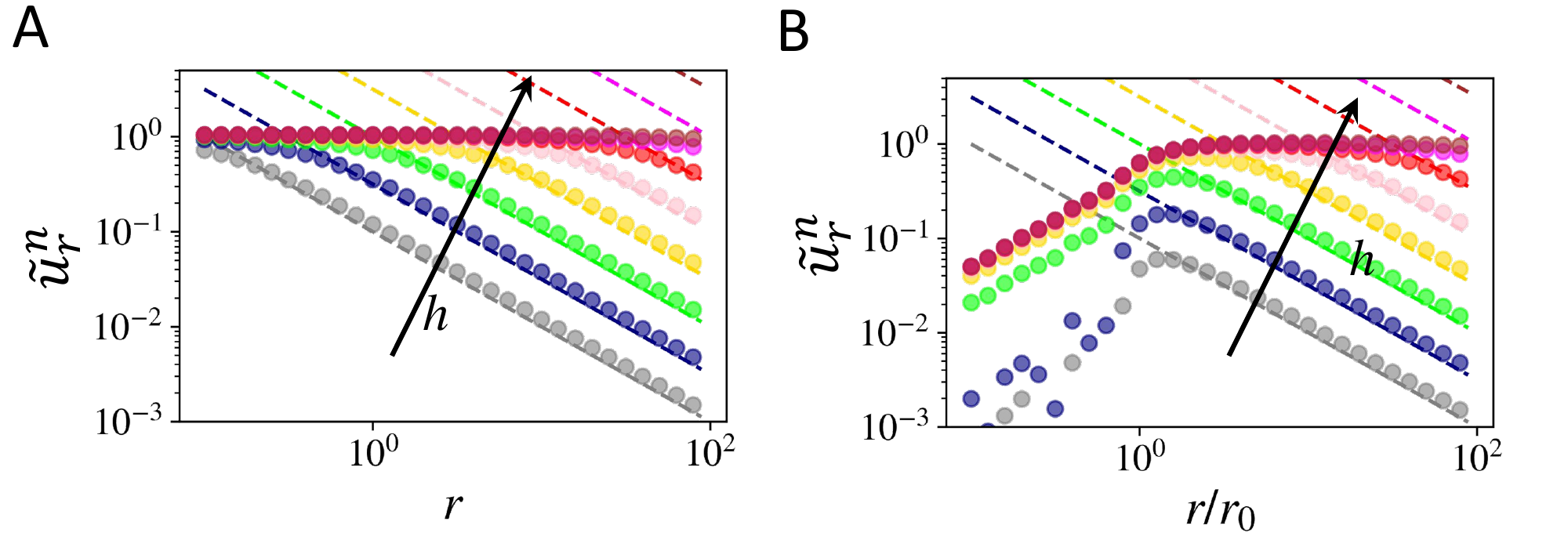}
\caption{The comparison of the non-slidable solution $\tilde{u}^n_r(r)$ (which is $u_r^n(r) $ normalized by $s\varepsilon/h\hat{G}$) between (A) with $r_0=0$ and (B) $r_0>0$. Different colors represent different thickness $h$ from 0.1 to $10^3$ in (A) and from $0.1r_0$ to $10^3r_0$ in (B).}
\label{fig:SI}
\end{figure}

\clearpage
\subsection*{Movies S1 to S3} 
\noindent Movie S1 bead displacement with weakly adhesive gel.\\
Movie S2 Radial displacement field measure at the surface of a weakly adhesive gel.\\
Movie S3 bead displacement with a gel rigidly bound to the glass.\\
\subsection*{Reference}
\cite{kawaue,RevModPhys.85.1143,Liu9944,Maruthamuthu4708,Liu9944,Khalilgharibi2019}
\end{document}